\documentclass[preprint,onecolumn,nofootinbib]{revtex4}
\pdfoutput=1
\usepackage[colorlinks=true,linkcolor=blue,urlcolor=blue,filecolor=black,citecolor=red,pdfstartview=FitV,pdftitle={},pdfsubject={},pdfkeywords={},pdfpagemode=None,bookmarksopen=true]{hyperref}
\usepackage{graphicx}
\usepackage{amsmath}
\usepackage{amsfonts}
\usepackage{amssymb,ulem}
\usepackage{color}%
\usepackage{tikz}
\usepackage{dcolumn}
\usepackage{xcolor}
\usepackage{xcolor}

\begin{document}
\title{
  Mixed-state Entanglement for AdS Born-Infeld Theory
}
\author{Peng Liu $^{1}$}
\email{phylp@email.jnu.edu.cn}
\author{Zhe Yang $^{1}$}
\email{yzar55@stu2021.jnu.edu.cn}
\author{Chao Niu $^{1}$}
\email{niuchaophy@gmail.com}
\author{Cheng-Yong Zhang $^{1}$}
\email{zhangcy@email.jnu.edu.cn}
\author{Jian-Pin Wu $^{2}$}
\email{jianpinwu@yzu.edu.cn}
\thanks{corresponding author}
\affiliation{
  $^1$ Department of Physics and Siyuan Laboratory, Jinan University, Guangzhou 510632, China\\
  $^2$ Center for Gravitation and Cosmology, College of Physical Science and Technology, Yangzhou University, Yangzhou 225009, China
}

\begin{abstract}

  We study the mixed-state entanglement for AdS Born-Infeld (BI) theory. We calculate the mixed-state entanglement and investigate the relationship between it and the system parameters. We find that the holographic entanglement entropy (HEE) and mutual information (MI) exhibit monotonically increasing and decreasing behavior with BI factor $b$. However, the entanglement wedge cross-section (EWCS) exhibits a very rich set of phenomena about system parameters. EWCS always increases with $b$ when $b$ is small and then monotonically decreases with $b$. These behaviors suggest that increasing the BI factor, which is essentially enhancing the coupling between the background geometry and the transport properties can always enhance the EWCS. The coupling between the entanglement and the transport behaviors has also been studied in condensed matter theories and is important to construct a stable quantum circuit. We also provide analytical understanding of the above phenomenon. Furthermore, we have tested two additional BI-like models and find the universality of these results, suggesting the crucial role of the BI term in governing the interplay between nonlinear electromagnetic effects and entanglement.
  
\end{abstract}
\maketitle
\tableofcontents

\section{Introduction}
\label{sec:introduction}

Quantum entanglement is the most distinguishing characteristic between quantum and classical systems. Holographic gravity, condensed matter theory, quantum information, and other areas have recently overlapped with each other on quantum entanglement. Numerous quantum entanglement measurements have been discovered to be capable of diagnosing the quantum phase transition of strong correlation systems and the topological quantum phase transitions, as well as playing a key role in the emergence of spacetime \cite{Osterloh:2002na, Amico:2007ag, Wen:2006topo, Kitaev:2006topo, Ryu:2006bv, Hubeny:2007xt, Lewkowycz:2013nqa, Dong:2016hjy}.

There are numerous types of quantum entanglement measurements, including entanglement entropy (EE), mutual information (MI), R\'enyi entanglement entropy, and negativity. Among these quantum entanglement measurements, EE is commonly considered a useful measure of pure state entanglement. However, EE is not applicable to measure the more common mixed-state entanglement. To measure mixed-state entanglement, numerous new entanglement measurements, such as the entanglement of purification, non-negativity, and the entanglement of formation, have been proposed \cite{vidal:2002, Horodecki:2009review}. On the other hand, calculating the mixed-state entanglement measures is extremely difficult.

Gauge/gravity duality is a powerful tool for analyzing strongly correlated systems because it connects entanglement-related physical quantities to geometric objects in dual gravity systems. In the dual gravity system, the holographic entanglement entropy (HEE) connects the EE of a subregion on the boundary with the area of the minimum surface \cite{Ryu:2006bv}. HEE has been demonstrated to be able to detect quantum phase transitions and thermodynamic phase transitions \cite{Nishioka:2006gr, Klebanov:2007ws, Pakman:2008ui, Zhang:2016rcm, Zeng:2016fsb}. Recently, the R\'enyi entropy was proposed to be proportional to the minimal area of cosmic branes \cite{Dong:2016fnf}. Moreover, the butterfly effect that reflects the dynamic properties of quantum systems, has been extensively studied in holographic theories \cite{Shenker:2013pqa, Sekino:2008he, Maldacena:2015waa, Donos:2012js, Blake:2016wvh, Blake:2016sud, Ling:2016ibq, Ling:2016wuy, Wu:2017mdl, Liu:2019npm}. In addition, the holographic duality of quantum complexity, a new information-related quantity from the EE, was also proposed \cite{Brown:2015lvg, Brown:2015bva, Chapman:2016hwi, Ling:2018xpc, Chen:2018mcc, Yang:2019gce, Ling:2019ien}. More recently, the EWCS was associated with the area of the minimum cross-section of the entanglement wedge \cite{Takayanagi:2017knl, Nguyen:2017yqw}. The geometric prescription of EWCS provides a novel and powerful tool for studying the mixed-state entanglement in holographic theories \cite{Chen:2021bjt, Cheng:2021hbw, Fu:2020oep, Gong:2020pse, Ling:2021vxe, Liu:2020blk, Liu:2021rks, Li:2021rff, Huang:2019zph}.

Among all the models in holographic theories, the Born-Infeld (BI) model is a special class of models for nonlinear electromagnetic field theories. It was first proposed to eliminate the divergent self-energy of the Maxwell theory. Later, it was found that the BI theory can be naturally derived from the string theory under the low-energy approximation. The BI model under the holographic theories can be dual to the quantum chromodynamics (QCD) systems \cite{Kundu:2013eba, Kundu:2019ull}, and some condensed matter systems with novel transport behaviors, such as the quantum liquid \cite{Karch:2009zz}, the Mott-insulator \cite{Baggioli:2016oju}, and the novel magneto-resistance phenomenon \cite{Kiritsis:2016cpm, Cremonini:2017qwq}, which is consistent with the experimental phenomenon in \cite{Hayers:2014zz, Hayes:2016}. Various properties of the BI model, such as its thermodynamic properties, transport properties \cite{Wu:2018zdc}, and the complexity \cite{Bakhtiarizadeh:2020kav}, have been extensively investigated. However, the question of how exactly the BI factor $b$, which embodies the nonlinearity of this nonlinear electromagnetic field theory, affects the properties of the system, especially the mixed-state entanglement properties, remains to be answered.

This paper focuses on the effect of the BI factor on two measures of mixed-state entanglement - MI and EWCS. When $b\to 0$, the background geometry is AdS-Schwarzschild solution, and the entanglement property of the system is decoupled from the transport property of the system; while for non-zero $b$, the transport behaviors can affect the entanglement property. Therefore, we interpret $b$ as the degree of correlation between the entanglement and transport properties of the metric when $b$ increases from zero. Remind also that the coupling between the transport properties and the entanglement is also an important topic in condensed matter field theory, and is crucial for the construction of a stable quantum circuit \cite{Mejia-Monasterio:2007,Hasan:2010,Stoermer:1999,Nandkishore:2015,nanowire:2011,plasom:2002,plasmonics:2013}. For $b\to \infty$, the system goes to the AdS-RN black brane system with a linear Maxwell field. Therefore, the range $b\in (0,\infty)$ represents the process that the Maxwell field turns on and converges to a linear Maxwell field case. Our main goal is to explore how BI factor $b$ affects the MI and EWCS. 

We organize this paper as follows: we introduce the holographic BI model in Sec. \ref{sec:phase_transition}, entanglement measures (HEE, MI, EWCS) and their holographic duality in Sec. \ref{subsec:info}. We discuss the properties of HEE, MI (\ref{sec:hee}) and EWCS (\ref{sec:eop_phenomena}) systematically. 
Subsequently, we examine two more BI-like models to further validate our results in \ref{sec:biaxmg}. Finally, we summarize in Sec. \ref{sec:discuss}.

\section{Holographic Born-Infeld Theory And Information-Related Quantities}\label{sec:alg}

First, we review the holographic BI model. Following that, we review the concepts of the HEE, MI, and EWCS with their holographic dual. Then, we elaborate upon our algorithms proposed to calculate minimum surfaces and minimum cross-sections.

\subsection{The AdS Born-Infeld Model}\label{sec:phase_transition}

The action of the 4-dimensional holographic BI model is,

\begin{equation}
  S = \int d ^ { 4 } x \sqrt { - g } \left[ \frac { R - 3 \Lambda } { 16 \pi G } + \frac { b ^ { 2 } } { 4 \pi G } \left( 1 - \sqrt { 1 + \frac { 2 \mathcal{F} } { b ^ { 2 } } } \right) \right].
\end{equation}
The parameter $b$ is the BI factor, $F_{\mu\nu}$ is the electromagnetic tensor and $\mathcal{F}=F^{\mu\nu}F_{\mu\nu}$. The cosmological constant $\Lambda = -\frac{3}{l^2}$ with $l$ the AdS radius. The equation of motion (EOM) of this model can be read as,
  \begin{equation}\label{eq:eom}
    \begin{aligned}
      \nabla_\mu\left(\frac{F^{\mu\nu}}{\sqrt{1+\frac{2\mathcal{F}}{b^2}}}\right)                                                                                                                         & =0,
      \\
      \mathcal{R}_{\mu\nu}-\mathcal{R} g_{\mu\nu}+\Lambda g_{\mu\nu}-2g_{\mu\nu}b^2\left(1-\sqrt{1+\frac{\mathcal{F}}{2b^2}}\right)-\frac{2 F_{\mu\rho}{F_\nu{^\rho}}}{\sqrt{1+\frac{\mathcal{F}}{2b^2}}} & =0.
    \end{aligned}
  \end{equation}

The solution of the BI theory is,
\begin{equation}
  \label{eq:fv1}
  d s ^ { 2 } = - f ( r ) d t ^ { 2 } + \frac { 1 } { f ( r ) } d r ^ { 2 } + r^2 h _ { i j } d x ^ { i } d x ^ { j },
\end{equation}
with
\begin{equation}
  f ( r ) = \frac { r ^ { 2 } } { l ^ { 2 } } - \frac { 2 M } { r } + \frac { 4 Q ^ { 2 } { } _ { 2 } F _ { 1 } \left( \frac { 1 } { 4 } , \frac { 1 } { 2 } ; \frac { 5 } { 4 } ; - \frac { Q ^ { 2 } } { b ^ { 2 } r ^ { 4 } } \right) } { 3 r ^ { 2 } } + \frac { 2 b ^ { 2 } r ^ { 2 } } { 3 } \left( 1 - \sqrt { \frac { Q ^ { 2 } } { b ^ { 2 } r ^ { 4 } } + 1 } \right),
\end{equation}
$Q$ is the electric charge and $M$ is the mass of the black brane. Also we have the gauge field $F_{\mu\nu}=\nabla_\mu A_\nu-\nabla_\nu A_\mu$ and $A_\mu=a(r)dt$,
  \begin{equation}
    a(r)=-\frac{\sqrt{2} Q \, _2F_1\left(\frac{1}{4},\frac{1}{2};\frac{5}{4};-\frac{2 Q^2}{r^4 b ^2}\right)}{r}.
  \end{equation}

Additionally, for $l^2 <0$ and $l^2>0$ the system is asymptotically dS and AdS, respectively. Here, we fix $l^2=1$ for concreteness. For $k=1,0,-1$ the $h_{ij}$ denotes a sphere, a Ricci flat surface, and a hyperbolic surface, respectively. Here, we focus on the planar case, i.e., $k=0$.

At the horizon $r=r_h$ we have $f(r_h)=0$, and hence we arrive at the ADM mass
\begin{equation}
  \label{eq:mass}
  M=\frac{4 l^2 Q^2 \, _2F_1\left(\frac{1}{4},\frac{1}{2};\frac{5}{4};-\frac{Q^2}{b^2 r_h^4}\right)-2 b^2 l^2 r_h^4 \sqrt{\frac{Q^2}{b^2 r_h^4}+1}+2 b^2 l^2 r_h^4+3 r_h^4}{6 l^2 r_h}.
\end{equation}
The Hawking temperature is,
\begin{equation}
  \label{eq:hawkingtemperature}
  T=\frac{r_h }{4 \pi }\left(3-2 b^2 \left(\sqrt{\frac{Q^2}{b^2 r_h^4}+1}-1\right)\right).
\end{equation}
The planar case is always thermodynamically stable \cite{Cai:2004eh}. Therefore, in this BI black brane system, there is no thermodynamic phase transition.

The system is invariant under the rescaling,
\begin{equation*}
  (t,1/r,x,y) \to \alpha (t,1/r,x,y),\;  Q \to Q/\alpha^2, \, T\to T/\alpha,\, r_h \to \alpha r_h .
\end{equation*}
Other parameters such as $b$ are dimensionless. Therefore, we can fix $r_h =1$. Here, we adopt $\sqrt{Q}$ as the scaling unit, consequently, we need to divide physical quantity with scaling dimension $[n]$ by $Q^{n/2}$.

\begin{figure}
  \centering
  \includegraphics[width =0.5\textwidth]{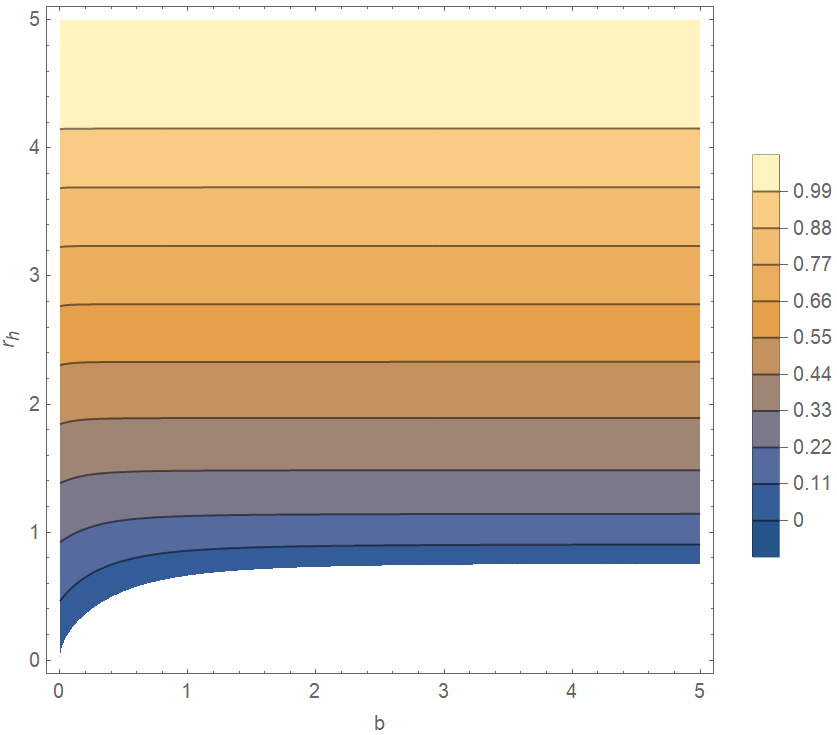}
  \caption{The contour plot of the Hawking temperature in the plane $(b,r_h)$, where the temperature is only positive in the shaded region.}
  \label{fig:contplot}
\end{figure}

For numerical convenience, we transform $r$ into $z \equiv r_h/r$ such that the semi-infinite domain $r\in (r_h,\infty)$ becomes a finite domain $z\in [0,1]$. Then the metric becomes,
\begin{equation}
  \label{eq:metric2}
  ds^2 = \frac{1}{z^2} \left( - h dt^2 + \frac{r_h^2 dz^2}{h}+r_h^2 dx^2+r_h^2 dy^2 \right),
\end{equation}
with
\begin{equation}
  \label{eq:hexp}
  \begin{aligned}
    h(z) \equiv & \frac{4}{3} Q^2 z^3 \left(z \, _2F_1\left(\frac{1}{4},\frac{1}{2};\frac{5}{4};-\frac{Q^2 z^4}{b^2}\right)-\, _2F_1\left(\frac{1}{4},\frac{1}{2};\frac{5}{4};-\frac{Q^2}{b^2}\right)\right) \\
                & -\frac{2}{3} b^2 \left(z^3 \left(1-\sqrt{\frac{Q^2}{b^2}+1}\right)+\sqrt{\frac{Q^2 z^4}{b^2}+1}-1\right)-z^3+1.
  \end{aligned}
\end{equation}
And the dimensionless Hawking temperature becomes,
\begin{equation}
  \label{eq:hawkingtemperaturescaled}
  T=\frac{b^2 \left(2-2 \sqrt{\frac{Q^2}{b^2}+1}\right)+3}{4 \pi  \sqrt{Q}}.
\end{equation}
From the dimensionless Hawking temperature \eqref{eq:hawkingtemperaturescaled} we can find that,
\begin{equation}
  \label{eq:limits}
  \lim_{Q\to 0} T \to \infty,\quad \lim_{Q\to \frac{\sqrt{12 b^2+9}}{2 b}} T \to 0.
\end{equation}
Also, we can find that,
\begin{equation}
  \label{eq:dqdt}
  \partial_Q T = 2 b \left(b-\sqrt{b^2+Q^2}\right)-3 \sqrt{\frac{Q^2}{b^2}+1}-2 Q^2 <0.
\end{equation}
Therefore, the quantity $Q$ is restricted to the range $[0,\frac{\sqrt{12 b^2+9}}{2 b}]$ and that the temperature $T$ decreases as $Q$ increases. This system is described by three variables $(T, \, b\,, r_h)$, with only two of them being independent. We have also observed that for any given value of $b$, the temperature $T$ always increases with increasing $r_h$, thus, the value of $r_h$ is uniquely determined by a given temperature $T$. This can be seen in the Fig. \ref{fig:contplot}. Therefore, we can simplify the system to a two-parameter system $(b,T)$.

When the parameter $b \to \infty$, the background solution of our system converges to the AdS-RN solution, and when $b\to 0$, it becomes the AdS-Schwarzschild solution. When $b$ is zero, there is an electromagnetic field present, but the background solution is still the AdS-Schwarzschild solution. This means that the entanglement-related physical quantities are not affected by the conductivity of the system. However, as $b$ increases, the electromagnetic fields starts to affect the background solution, and thus has an impact on the entanglement structure of the system. Therefore, we refer to increasing $b$ from zero to infinity as the process of turning on the coupling between the background and the conductivity, and finally resulting in an AdS-RN system.

It is worth noting that the relationship between conductivity and entanglement-related quantities is of great importance in condensed matter theories. 

  In most condensed matter systems, quantum entanglement shows an intricate connection to transport properties like thermal and electrical conductivity \cite{Mejia-Monasterio:2007,Hasan:2010,Stoermer:1999,Nandkishore:2015}. The underlying quantum coherence enables nonlocal correlations can directly manifest in the measurable transport. For instance, in quantum spin chains, the entanglement between distant spins allows energy transport despite the lack of individual particle motion \cite{Mejia-Monasterio:2007}. However, in certain important cases, the entanglement can become decoupled from the transport behavior. For example, in topological insulators, the robust edge state transport persists regardless of the bulk entanglement due to topological protection \cite{Hasan:2010}. This illustrates the irrelevance between entanglement and transport. Though counterintuitive, such irrelevance underscores the delicate nature of quantum effects.

Recent experiments have shown that entanglement between quantum dots can persist despite the influence of surface plasmon polariton (SPPs) transmission \cite{plasmonics:2013,plasmon:2004,plasom:2002}. These findings are crucial for the development of stable quantum circuits. Additionally, it has been found that at specific values of the inter-dot distance $d$ or detuning $\delta$, the two-quantum-dot system can be in a highly entangled state and be separate from the transmission of SPPs \cite{nanowire:2011}. However, when $d$ or $\delta$ deviate from these values, the entanglement of quantum dots becomes highly correlated with the transmission of SPPs. This suggests that decoupling of entanglement and transport can exist in real physical systems and can be characterized by certain parameters. Properly modeling the coupling and decoupling between entanglement and transport remains an open challenge. Advanced theoretical frameworks need to be developed that can transition between the two regimes.

Next, we will focus on how the entanglement-related physical quantities change as we vary the parameter $b$.

\subsection{Holographic information-related quantities}\label{subsec:info}

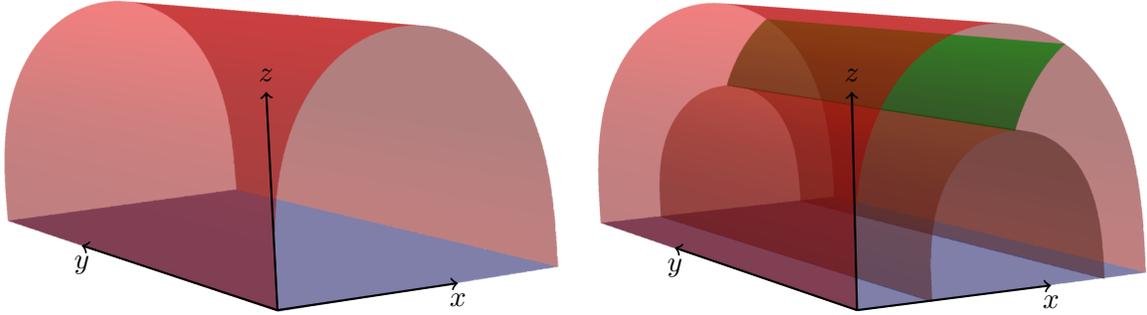
\begin{figure}
  \begin{tikzpicture}[scale=1]
    \node [above right] at (0,0) {\includegraphics[width=7.5cm]{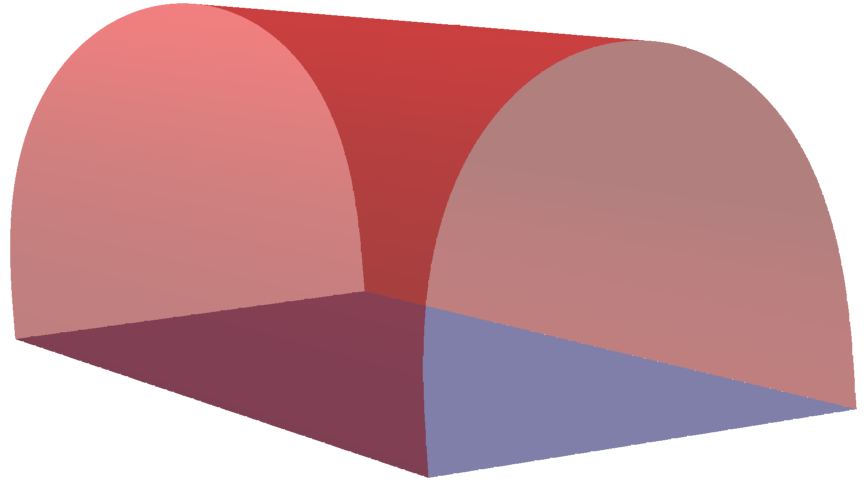}};
    \draw [right,->,thick] (3.85, 0.22) -- (6.25, 0.58) node[below] {$x$};
    \draw [right,->,thick] (3.85, 0.22) -- (1.25, 1.08) node[below] {$y$};
    \draw [right,->,thick] (3.85, 0.22) -- (3.7, 3.125) node[above] {$z$};
  \end{tikzpicture}
  \begin{tikzpicture}[scale=1]
    \node [above right] at (0,0) {\includegraphics[width=7.5cm]{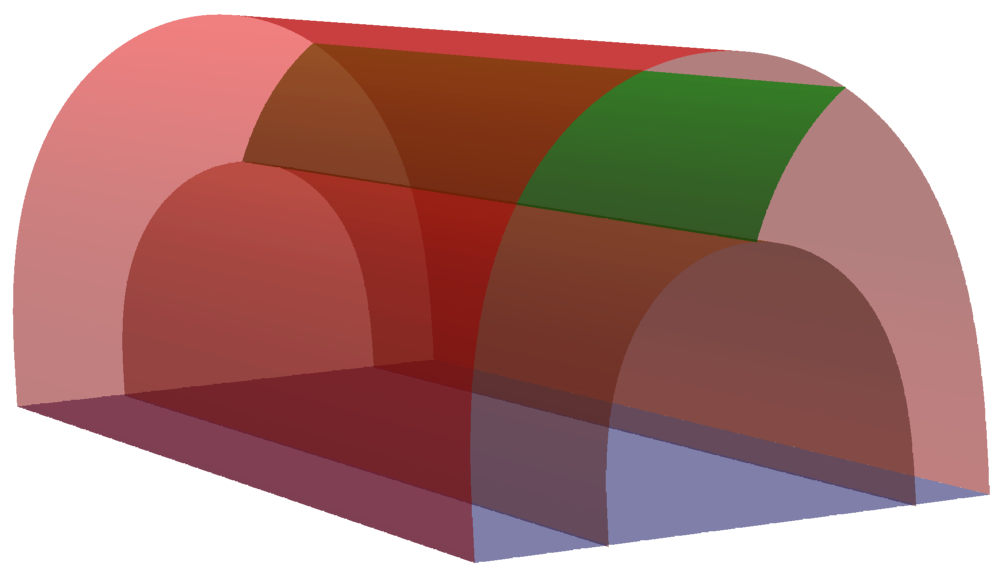}};
    \draw [right,->,thick] (3.67, 0.22) -- (6.25, 0.55) node[below] {$x$};
    \draw [right,->,thick] (3.67, 0.22) -- (1.25, 1.05) node[below] {$y$};
    \draw [right,->,thick] (3.67, 0.22) -- (3.6, 3.125) node[above] {$z$};
  \end{tikzpicture}
  \caption{The left plot: The minimum surface for a given width $w$. The right plot: The minimum cross-section (green surface) of the entanglement wedge.}
  \label{msd1}
\end{figure}

Entanglement is a fundamental and intriguing aspect of quantum mechanics. One way to quantify entanglement is through entanglement entropy (EE), which measures the degree of entanglement between a subset of a system and the rest of the system. Specifically, the entanglement entropy $S_A$ between subsets $A$ and $B$ of a system $A\cup B$ is defined as the von Neumann entropy in terms of the reduced density matrix $\rho_A$.
\begin{equation}\label{ee-von}
  S_{A} (|\psi\rangle) = - \text{Tr}\left[ \rho_{A} \log \rho_{A} \right],\quad \rho_{A} = \text{Tr}_{B} \left(|\psi\rangle\langle\psi|\right).
\end{equation}
It is easy to find that $S_A = S_B$ for pure states \cite{Chuang:2002book}. Holographic duality theory relates the holographic entanglement entropy (HEE) to the area of the minimum surface in dual gravity systems \cite{Ryu:2006bv} (see the left plot of Fig. \ref{msd1}).

EE is often used to measure the degree of entanglement in pure states, but it is not as effective in measuring mixed state entanglement. For example, even when subsystems $A$ and $B$ are not entangled, they can still have non-zero EE in a system composed of direct product of the density matrices of $\rho_A$ and $\rho_B$. This is because EE takes into account both quantum entanglement and classical correlation, so it does not always provide a accurate measure of the entanglement. As a result, other measures for mixed-state entanglement have been proposed in the literature \cite{vidal:2002,Horodecki:2009review}. The most direct mixed-state entanglement measure is MI. 

For the subsystem $A\cup C$ separated by $B$, the mutual information (MI) is defined as:
\begin{equation}\label{mi:def}
  I\left(A,B\right) := S\left(A\right) + S\left(B\right) - S\left(A\cup B\right),
\end{equation}
This measures the mixed-state entanglement between $A$ and $B$. It can be easily verified that $ I\left(A,B\right) =0 $ when $\rho_{AB} = \rho_{A} \otimes \rho_{B} $, therefore MI have the property that direct product states have zero entanglement. However, MI is not a perfect measure of mixed-state entanglement, as it is closely related to EE, and it's properties are sometimes dominated by EE or thermal entropy in certain situations. This indicates that other measures of mixed-state entanglement should be used.

The entanglement wedge cross-section (EWCS) has been associated with the duality of certain mixed-state entanglement measures, such as entanglement of purification, logarithmic negativity, and reflect entropy \cite{Kudler-Flam:2018qjo,Kusuki:2019zsp,Dutta:2019gen}. Takayanagi proposed that EWCS $ E_{W}\left(\rho_{AB}\right) $ is the area of the minimum cross-section $ \Sigma_{AB} $ in connected entanglement wedge \cite{Takayanagi:2017knl}, {\it i.e.} (see the right plot in Fig. \ref{msd1}),
\begin{equation}\label{heop:def}
  E_{W}\left(\rho_{AB}\right) = \min_{\Sigma_{AB}} \left( \frac{\text{Area} \left(\Sigma_{AB}\right)}{4G_{N}}\right) .
\end{equation}
It is important to note that if the entanglement wedge is disconnected, meaning the minimum cross-section does not exist, the EWCS will be zero, it corresponds to cases with vanishing MI. Additionally, the EWCS also satisfies some important inequalities as its quantum information counterparts \cite{Takayanagi:2017knl,Bao:2017nhh}

Next, we present the algorithm for obtaining the minimum surfaces and EWCS.

\subsection{Computations of holographic geometric quantities}\label{eopcomp}

We examine the EWCS of an infinite strip with a homogeneous background for numerical simplicity.
For a generic homogeneous background
\begin{equation}\label{genbg}
  ds^{2} = {g_{tt}} dt^2 + g_{zz}dz^2 + g_{xx}dx^2 + g_{yy} dy^2,
\end{equation}
where $z=0$ represents the boundary of the asymptotic AdS spacetime. The left plot in Fig. \ref{fig:cartoon4eop} is a visual representation of the minimum surface for an infinite strip along the $y$-axis. Since the background is homogeneous, all metric components $g_{\mu\nu}$ only depend on the coordinate $z$.

\subsubsection{The minimum surface}

The minimum surface near the AdS boundary is perpendicular to the boundary, making the spatial direction $x$ an unsuitable parameter for finding the minimum surface. Ref. \cite{Ling:2019tbi} adopted the angle $\theta$ with $\tan\theta = z/x$, as the parameter for the minimum surface (see Fig. \ref{fig:cartoon4eop}). Using this method, we can parametrize a surface as $(x(\theta),z(\theta))$ with area $A$ given by
\begin{equation}\label{eq:lag}
  A = 2\int_{0}^{\pi/2} \sqrt{x'(\theta )^2 g_{xx} g_{yy}+z'(\theta )^2 g_{yy} g_{zz}} d\theta.
\end{equation}
The resultant equations of motion read,
\begin{equation}\label{eq:eom2}
  \begin{aligned}
    x'(\theta ) z'(\theta )^2 \left(\frac{g_{ xx }'}{2 g_{ xx }}+\frac{g_{ yy }'}{g_{ yy }}-\frac{g_{ zz }'}{2 g_{ zz }}\right)+\frac{x'(\theta )^3 \left(g_{ yy } g_{ xx }'+g_{ xx } g_{ yy }'\right)}{2 g_{ xx } g_{ zz }}+x''(\theta ) z'(\theta )-x'(\theta ) z''(\theta ) & =0,  \\
    z(\theta) - \tan(\theta) x(\theta)                                                                                                                                                                                                                                         & =0.
  \end{aligned}
\end{equation}
where $g_{\#\#}'\equiv g_{\#\#}'(z)$. The boundary conditions are,
\begin{equation}\label{eq:bcs}
  z(0)=0,\quad x(0)=w,\quad z'(\pi/2)=0,\quad x(\pi/2)=0,
\end{equation}
where $w$ is the width of the strip.

\subsubsection{The EWCS}

Given a biparty subsystem with minimum surfaces $C_1(\theta_1),\,C_2(\theta_2)$, we solve the minimum surface $C_{p_1,p_2}$ connecting $p_1 \in C_1$ and $p_2\in C_2$. We parametrize $C_{p_1,p_2}$ with $z$, then the area of $C_{p_1,p_2}$ reads,
\begin{equation}\label{eq:zpara}
  A = \int_{C_{p_1,p_2}} \sqrt{ g_{xx} g_{yy} x'(z)^2 + g_{xx} g_{zz} } dz.
\end{equation}
The resultant equation of motion becomes,
\begin{equation}\label{eq:zparaeom}
  x'(z)^3 \left(\frac{g_{ xx } g_{ yy }'}{2 g_{ yy } g_{ zz }}+\frac{g_{ xx }'}{2 g_{ zz }}\right)+x'(z) \left(\frac{g_{ xx }'}{g_{ xx }}+\frac{g_{ yy }'}{2 g_{ yy }}-\frac{g_{ zz }'}{2 g_{ zz }}\right)+x''(z) =0,
\end{equation}
with boundary conditions,
\begin{equation}\label{eq:zparabcs}
  x(z(\theta_1)) = x(\theta_1),\quad x(z(\theta_2)) = x(\theta_2).
\end{equation}
To obtain the EWCS, we need to locate the global minimum of the minimum surfaces connecting $C_1(\theta_1),\,C_2(\theta_2)$, i.e., the minimum cross-section.

\begin{figure}
  \centering
  \includegraphics[width = \textwidth]{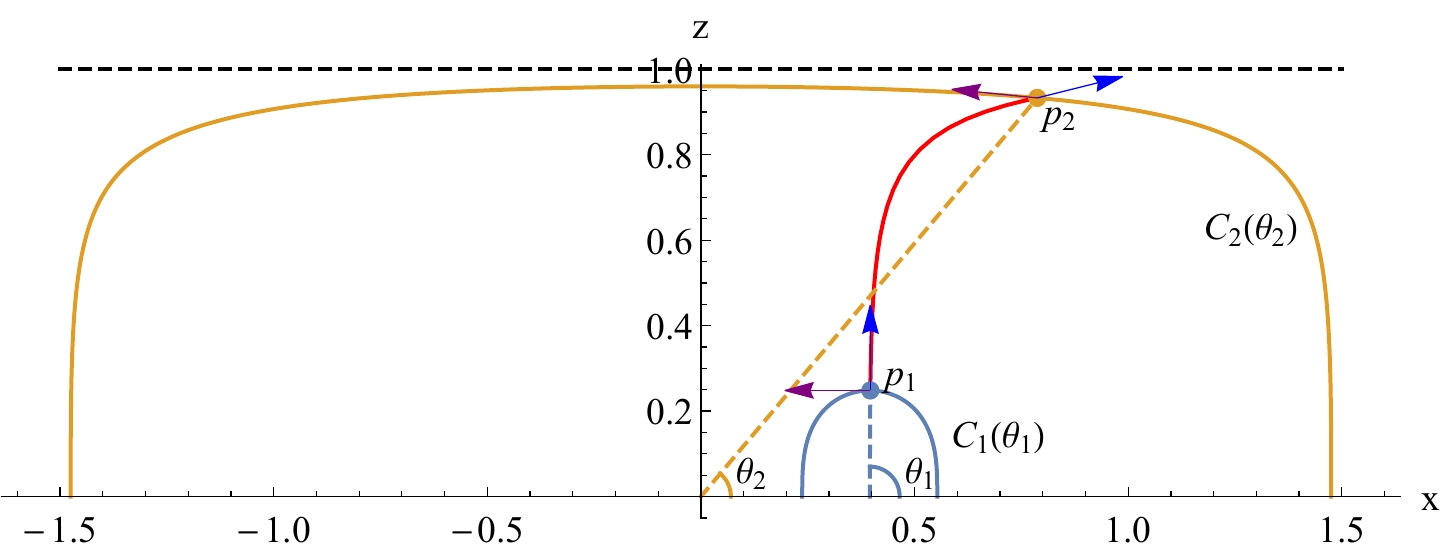}
  \caption{
  The demonstration of the EWCS. The $p_1$ and $p_2$ are the intersection points of the minimum surface connecting those two minimum surfaces. The solid blue curve (parametrized with $\theta_1$) and solid orange curve (parametrized with $\theta_2$) are minimum surfaces. The thick red curve is the minimum surface connecting $p_1$ and $p_2$. The blue arrows at the $p_1$ and $p_2$ are the tangent vector $\left.\left(\frac{\partial}{\partial z}\right)^a\right|_{p_1}$ and $\left.\left(\frac{\partial}{\partial z}\right)^a\right|_{p_2}$ along the $C_{p_1,p_2}$, while the purple arrows are the tangent vectors $\left.\left(\frac{\partial}{\partial \theta_1}\right)^a\right|_{p_1}$ and $\left.\left(\frac{\partial}{\partial \theta_2}\right)^a\right|_{p_2}$ along $C_1,\,C_2$, respectively. The dark dashed horizontal line is the horizon.
  }
  \label{fig:cartoon4eop}
\end{figure}

Finding the minimum cross-section is a challenging task as it involves searching through a two-dimensional parameter space $(\theta_1,\theta_2)$. However, it can be noted that the globally minimum cross-section must be perpendicular to the minimum surfaces at the point of intersection. This observation serves as a local constraint, which can greatly speed up the search process. We demonstrate the methods of solving the EWCS in Fig. \ref{fig:cartoon4eop}.
For numerical stability, it is better to implement the perpendicular conditions with normalized vectors as,
\begin{equation}\label{eq:perpend2}
  \begin{aligned}
    Q_1(\theta_1,\theta_2) & \equiv \left.\frac{g_{ab} \left(\frac{\partial}{\partial z}\right)^a \left(\frac{\partial}{\partial \theta_1}\right)^b}{\sqrt{g_{cd} \left(\frac{\partial}{\partial z}\right)^c \left(\frac{\partial}{\partial z}\right)^d} \sqrt{g_{mn} \left(\frac{\partial}{\partial \theta_1}\right)^m \left(\frac{\partial}{\partial \theta_1}\right)^n}}\right|_{p_1} = 0,  \\
    Q_2(\theta_1,\theta_2) & \equiv \left.\frac{g_{ab} \left(\frac{\partial}{\partial z}\right)^a \left(\frac{\partial}{\partial \theta_2}\right)^b}{\sqrt{g_{cd} \left(\frac{\partial}{\partial z}\right)^c \left(\frac{\partial}{\partial z}\right)^d} \sqrt{g_{mn} \left(\frac{\partial}{\partial \theta_2}\right)^m \left(\frac{\partial}{\partial \theta_2}\right)^n}}\right|_{p_2} = 0.
  \end{aligned}
\end{equation}
Note that $Q_1$ and $Q_2$ are both functions of the $\theta_1$ and $\theta_2$. Now, the search of the EWCS is equivalent to finding the minimum surface ending at $(\theta_1,\theta_2)$ where \eqref{eq:perpend2} is satisfied.

To determine the correct EWCS, we first select an initial seed $(\theta_1,\theta_2)$ and use the Newton iterative method to obtain feedback $(\delta \theta_1,\delta \theta_2)$. By repeating this process, we can find the minimum cross-section, which is the EWCS. It is crucial to carefully choose the initial values of $(\theta_1,, \theta_2)$ for the iterations to converge. The numerical reliability is ensured by the convergence of results when using different initial values or increasing the density of discretization. For more technical details, refer to reference \cite{Boyd:2001}.

Using the techniques outlined above, we will now examine mixed-state entanglement measures for the BI model. Additionally, we will examine the correlation between the BI factor $b$ and information-related quantities.

\section{The Holographic Entanglement Entropy And The Holographic Mutual Information}\label{sec:hee}

\begin{figure}
  \centering
  \includegraphics[width=0.45\textwidth]{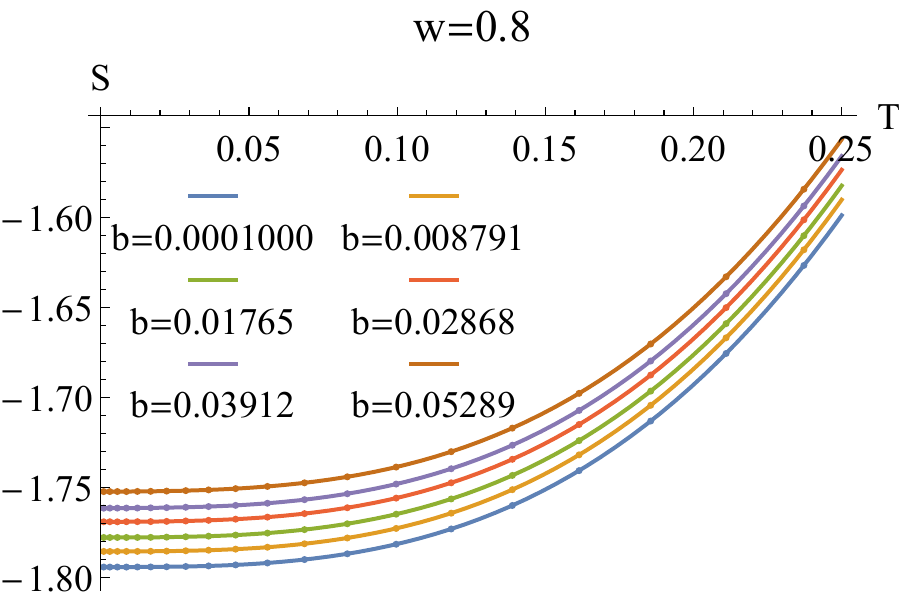}\quad
  \includegraphics[width=0.45\textwidth]{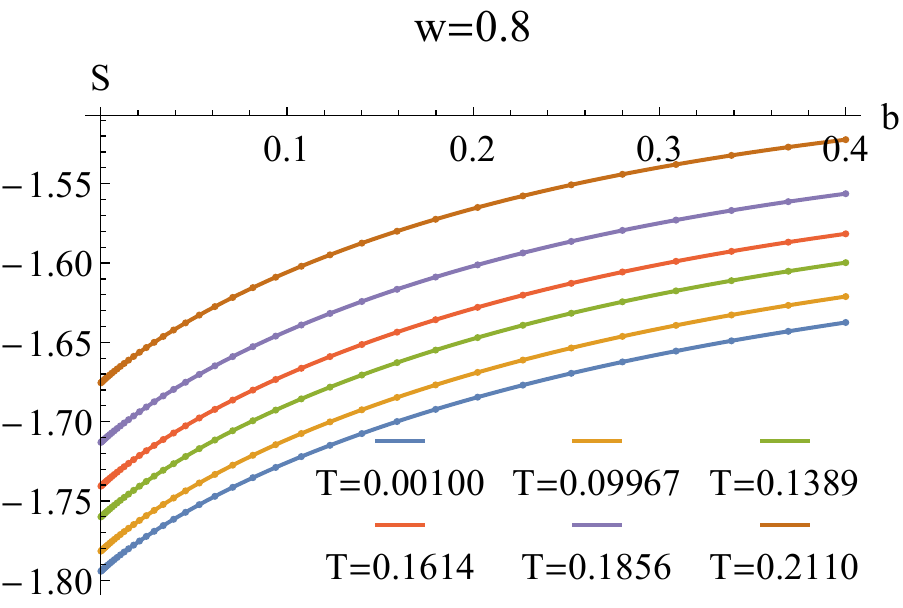}
  \caption{HEE vs $T$ and $b$ at width $w=0.8$, respectively.}
  \label{fig:heetest}
\end{figure}

We begin by examining the relationship between HEE, system parameters $b$ and $T$. As shown in Fig. \ref{fig:heetest}, HEE, represented by $S$, increases monotonically with both $b$ and $T$, but their rate of increase is different. Initially, $S$ increases slowly with $T$ and its growth rate with $T$ becomes more pronounced as $T$ increases. On the other hand, $S$ increases quickly with $b$ at first and then slows down as $b$ decreases. 
Next, we explain the behavior of $ S $ with $ b $ and $ T $, respectively.

On the gravity side, the HEE is related to the area of the minimum surface. Especially, when the horizon radius of the black brane increases, the minimum surface tends to be closer to the horizon of the black brane, which makes the thermodynamic entropy dominate the behavior of the HEE. Therefore, the growth of HEE with $ T $ as well as $ b $, can be understood from the relation between $r_h$ and $T$ or $b$.
According to \eqref{eq:hawkingtemperature} we can deduce that $r_h$ increases with increasing temperature and $b$, this can be seen by taking the derivative of $r_h$ with respect to $T$ and $b$. The results are,
\begin{equation}\label{eq:dtdrhdb}
  \begin{aligned}
    \partial_T r_h & =\frac{r_h^4 \sqrt{\frac{Q^2}{b^2 r_h^4}+1}}{r_h^4 \left(2 b^2 \left(\sqrt{\frac{Q^2}{b^2 r_h^4}+1}-1\right)+3 \sqrt{\frac{Q^2}{b^2 r_h^4}+1}\right)+2 Q^2},                                                             \\
    \partial_b r_h & =\frac{2 r_h \left(Q^2-2 b^2 r_h^4 \left(\sqrt{\frac{Q^2}{b^2 r_h^4}+1}-1\right)\right)}{b \left(r_h^4 \left(2 b^2 \left(\sqrt{\frac{Q^2}{b^2 r_h^4}+1}-1\right)+3 \sqrt{\frac{Q^2}{b^2 r_h^4}+1}\right)+2 Q^2\right)}.
  \end{aligned}
\end{equation}
From the above equation, it is clear that $\partial_Tr_h$ is always positive, indicating that $r_h$ increases as $T$ increases. However, $\partial_br_h$ can be positive or negative, depending on the specific parameter range. Further examination shows that $\partial_br_h$ is always greater than zero when $r_h$ is relatively large. This means that $r_h$ increases with $b$ when $r_h$ is large, or when the minimum surface is closer to the horizon of the black brane. When $b$ is relatively large, the system is approximately the AdS-RN system. The argument presented in \cite{Liu:2019qje} can be applied to prove that $\partial_T S>0$. Furthermore, for small subregions, it can be inferred from the equations in \cite{Liu:2019qje} that $\partial_T S$ is close to $0$, which explains the flat behavior of $S$ along $T$ for small temperatures.

\begin{figure}
  \centering
  \includegraphics[height=0.4\textwidth]{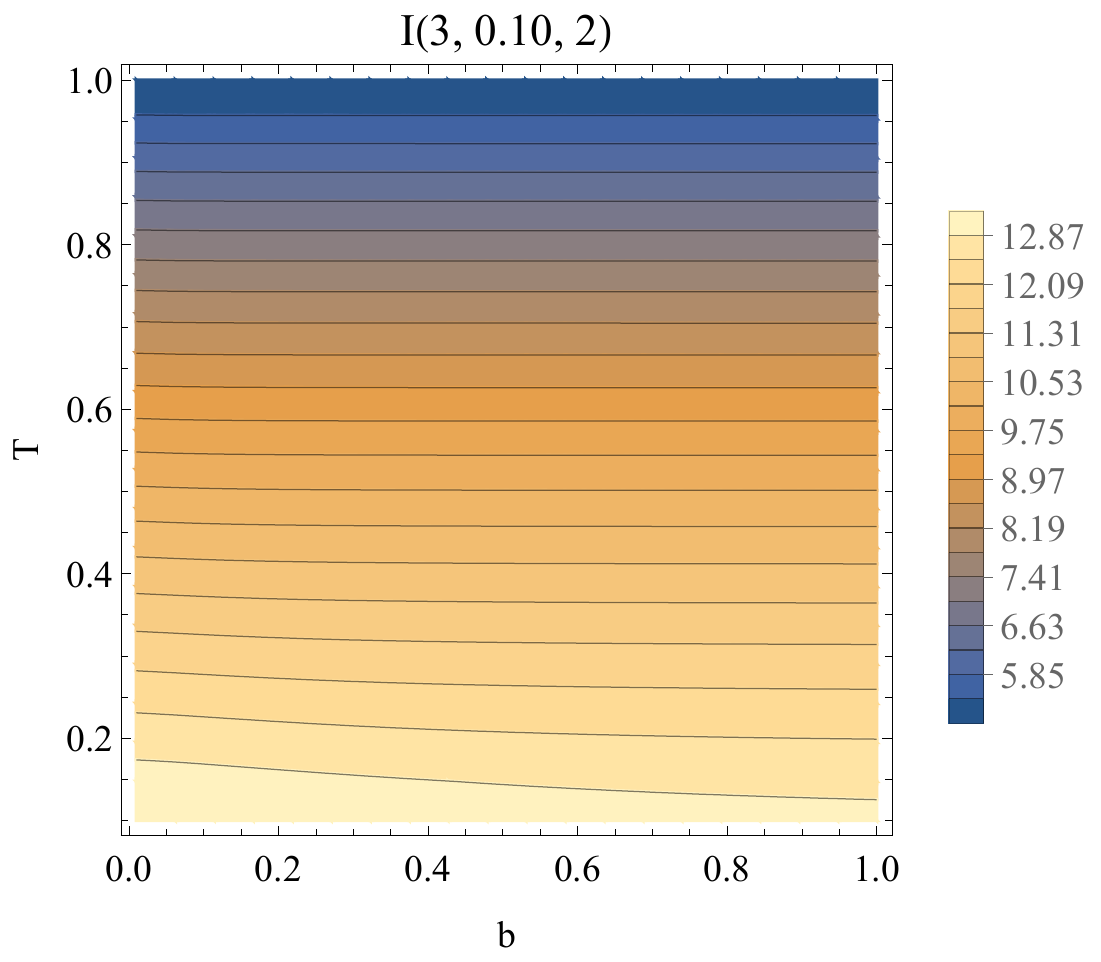}
  \includegraphics[height=0.4\textwidth]{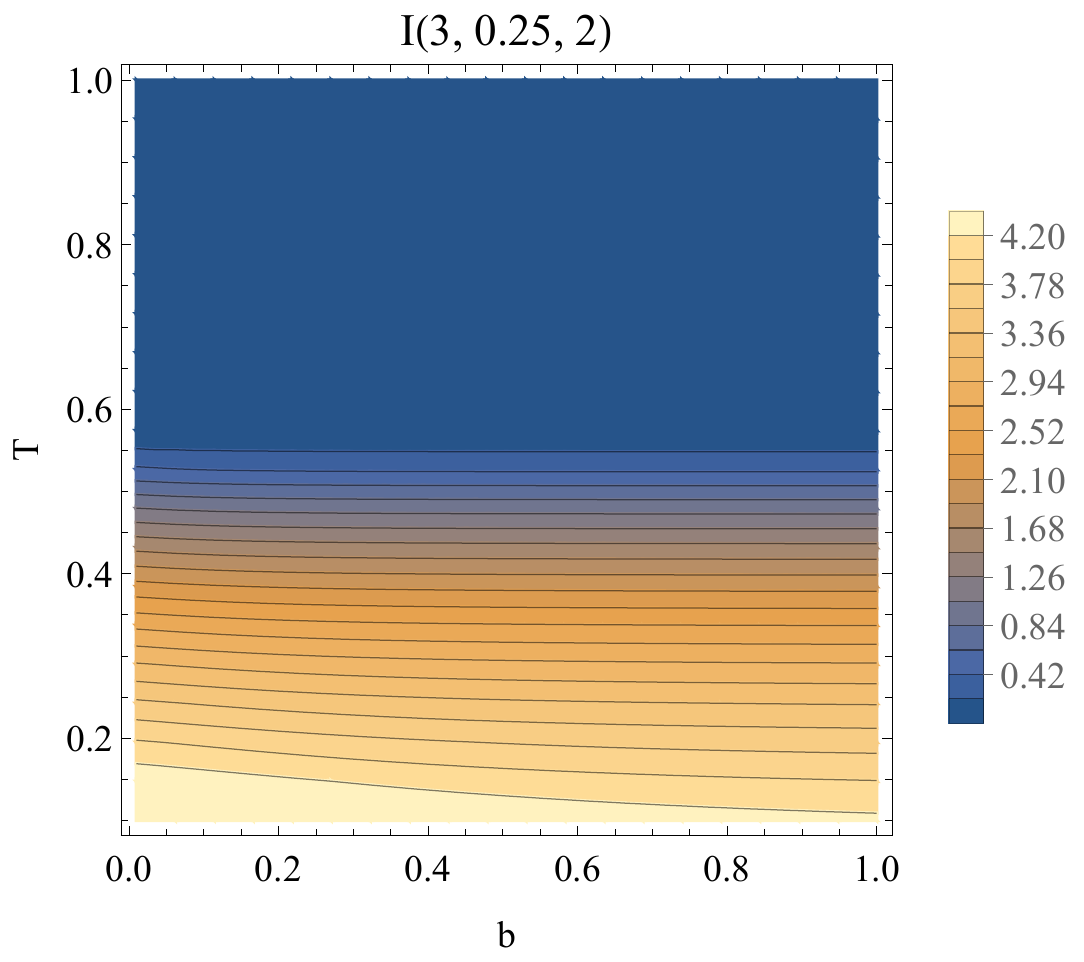}
  \caption{
    MI as a function of $b$ and $T$ for different configurations.
  }
  \label{fig:mi}
\end{figure}

\begin{figure}
  \centering
  \includegraphics[width=0.45\textwidth]{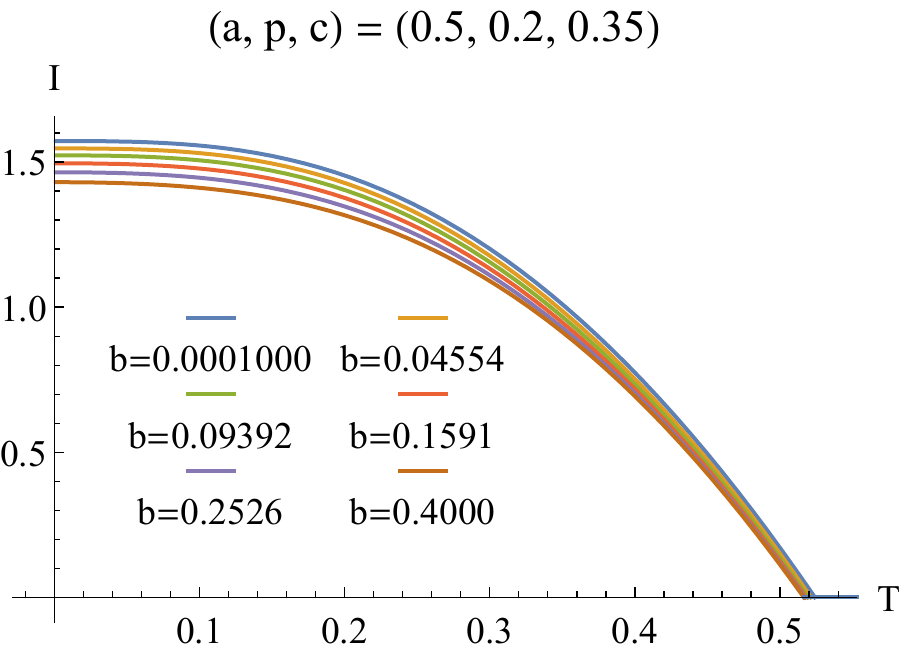}\quad
  \includegraphics[width=0.45\textwidth]{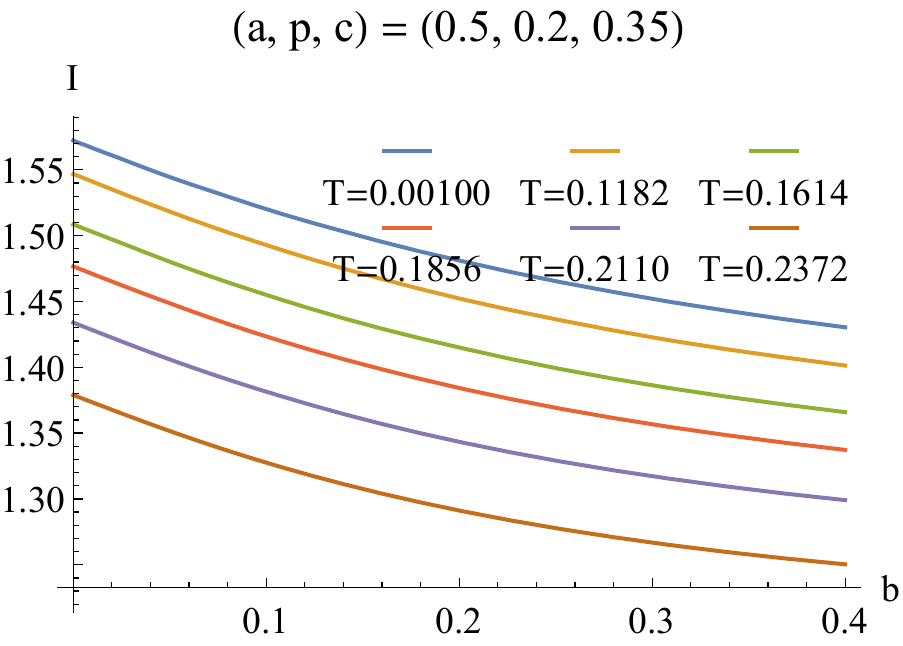}
  \caption{MI as a function of $b$ and $T$ for different configurations.}
  \label{fig:mic}
\end{figure}

After studying HEE, we proceed to investigate the behavior of MI with $T$ and $b$. In the BI model, the configurations for MI and EWCS are subsystems composed of $a$ and $b$ separated by region $p$. As seen in Fig. \ref{fig:mi}, MI decreases with increasing temperature and $b$. This is in contrast to the behavior of HEE. Moreover, it is worth noting that MI can decrease to zero, which is an indication of a disentanglement phase transition. We have also plotted the MI for smaller configurations (see Fig. \ref{fig:mic}), and the qualitative phenomena remain the same.

\begin{figure}
  \centering
  \includegraphics[width=0.45\textwidth]{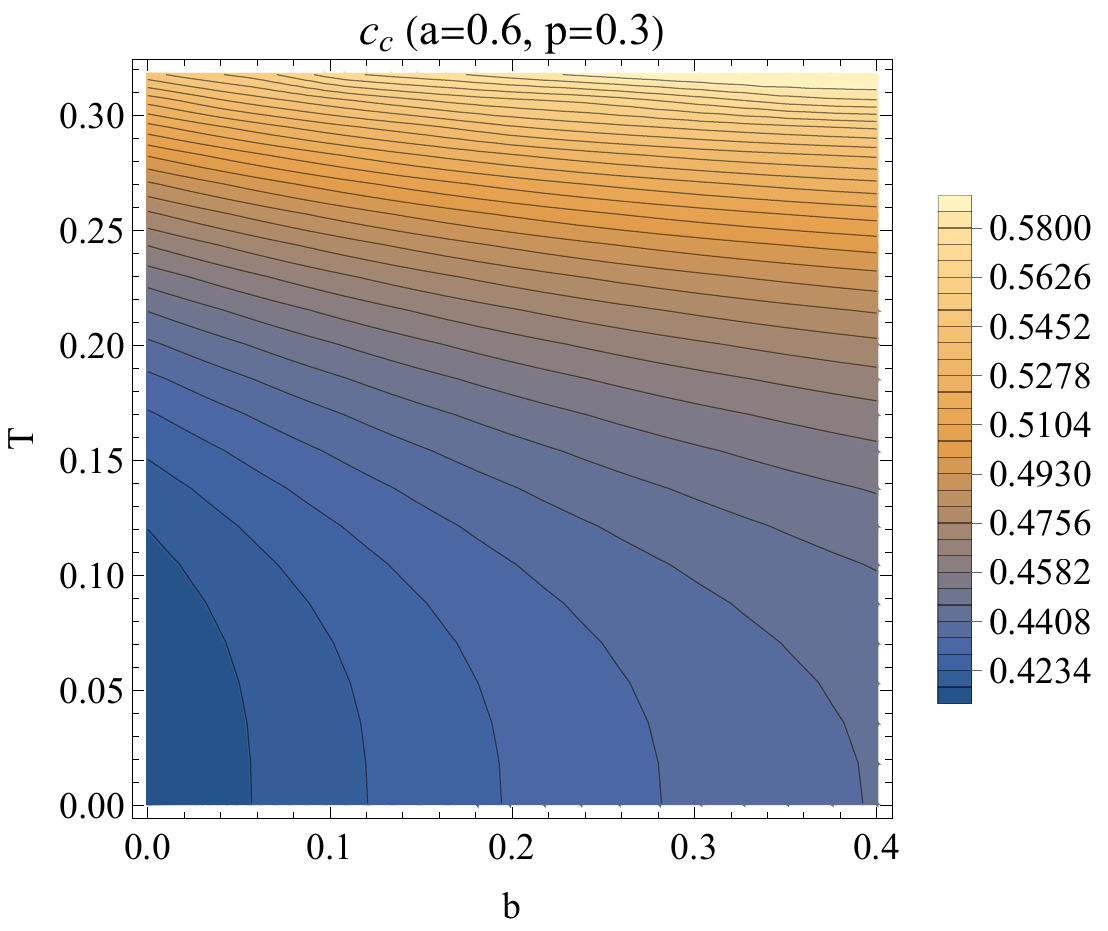}\quad
  \includegraphics[width=0.45\textwidth]{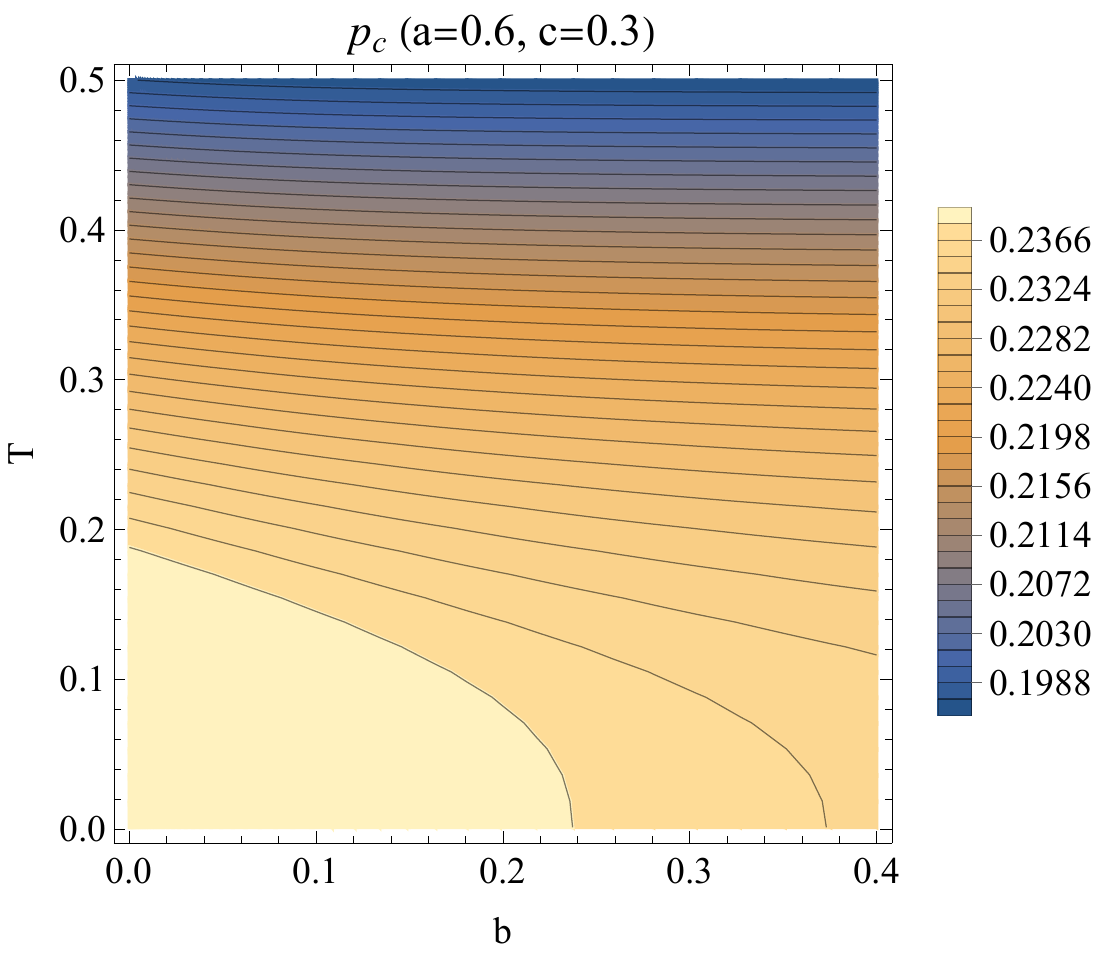}
  \caption{Critical configurations of $c_c$ and $p_c$.}
  \label{fig:cric}
\end{figure}

As the subsystem $c$ and the separation $p$ change, the system undergoes a disentangling phase transition, at which point the entanglement of two subsystems $a$ and $c$ vanishes. The critical value of subsystem $c_c$ and separation $p_c$ are shown in Fig. \ref{fig:cric}. The left plot of Fig. \ref{fig:cric} shows that the critical value of subsystem $c_c$ increases with $b$ and $T$; however, the right plot of Fig. \ref{fig:cric} shows that the critical value of the separation $p_c$ decreases with $b$ and $T$. This is as expected since increasing the temperature or $b$ will tends to destroy the entanglement, resulting in a larger subregion $c_c$ or a smaller separation $p_c$.

  In summary, both HEE and MI exhibit monotonic behavior with respect to the BI factor $b$, but in opposite manner - HEE increases while MI decreases. From the gravity side, the thermal contribution to HEE and MI dominates as $r_h$ increases with $b$, which leads to the monotonic dependence on $b$. From a perspective rooted in the AdS/CFT correspondence, the thermal effects dominate the entanglement behavior observed on the boundary CFT. However, EWCS, being minimum cross-sections anchored on the minimum surfaces, may exhibit different behavior from the black hole thermal physics. Investigating EWCS could unveil new perspectives on mixed state entanglement.

Next, we explore the mixed-state entanglement through the EWCS.

\section{The Holographic Entanglement Wedge Cross-Section}\label{sec:eop_phenomena}

\begin{figure}
  \centering
  \includegraphics[width=0.8\textwidth]{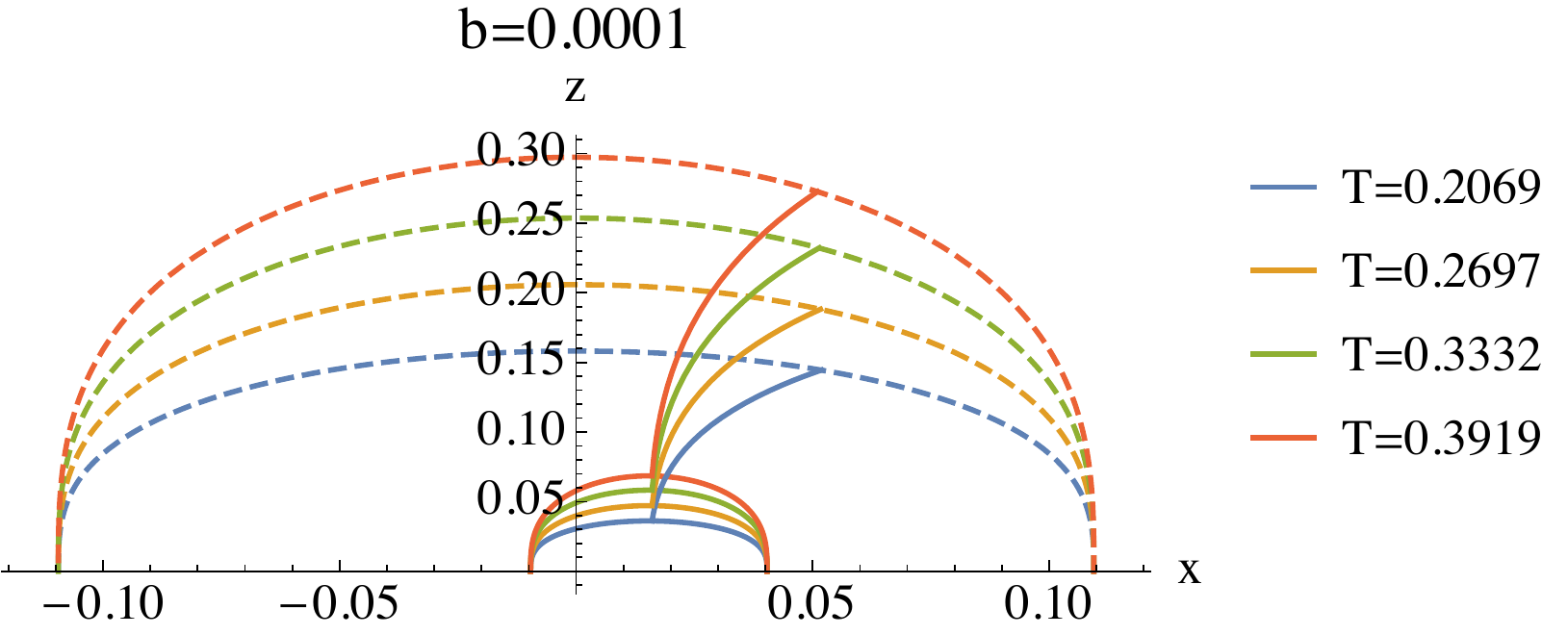}
  \caption{The illustration of EWCS. At the same configuration $(a,p,c) = (0.1, 0.05, 0.06925)$ we see that the minimum surface becomes flatter when decreasing the temperature. Meanwhile, the minimum cross-section always ends at the point near the tops of the inner minimum surface, while ends at the point away from the tops of the outer minimum surface.}
  \label{fig:eopshow}
\end{figure}

In Fig. \ref{fig:eopshow}, we present the minimum surfaces and the corresponding minimum cross-sections. It can be observed that the minimum surface is flatter when the temperature is lower. This is due to the fact that the coordinate $z$ is related to the horizon radius $r_h$, and at lower temperatures, a small $r_h$ will rescale $z$ to $z r_h$, resulting in a flatter minimum surface. This makes it challenging to obtain precise enough solutions for the minimum surface since the flat case is more singular in the $\theta$ coordinate. To overcome this issue, we redefine the angle as $z = \eta x \tan (\theta)$, where $\eta$ is a number related to the temperature. Only with this technique, we can achieve precise enough solutions.

\begin{figure}
  \centering
  \includegraphics[width=0.45\textwidth]{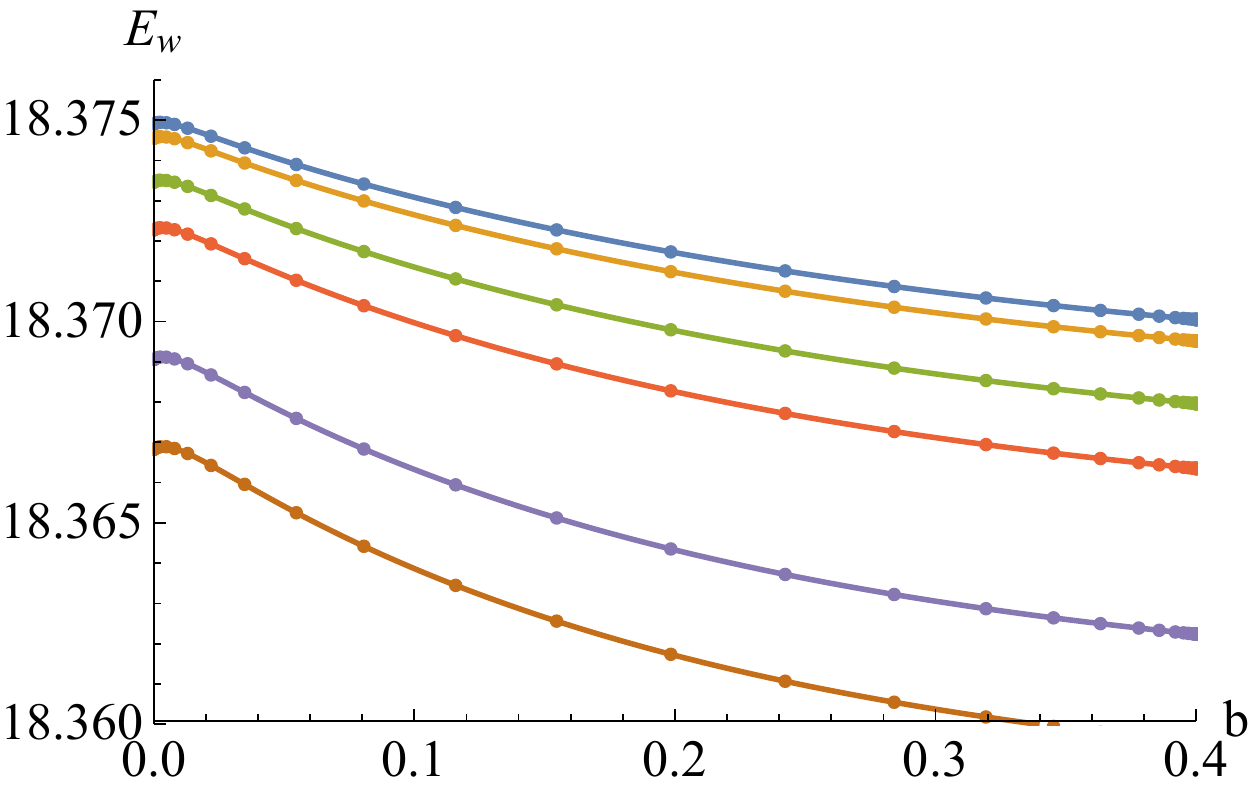}
  \includegraphics[width=0.45\textwidth]{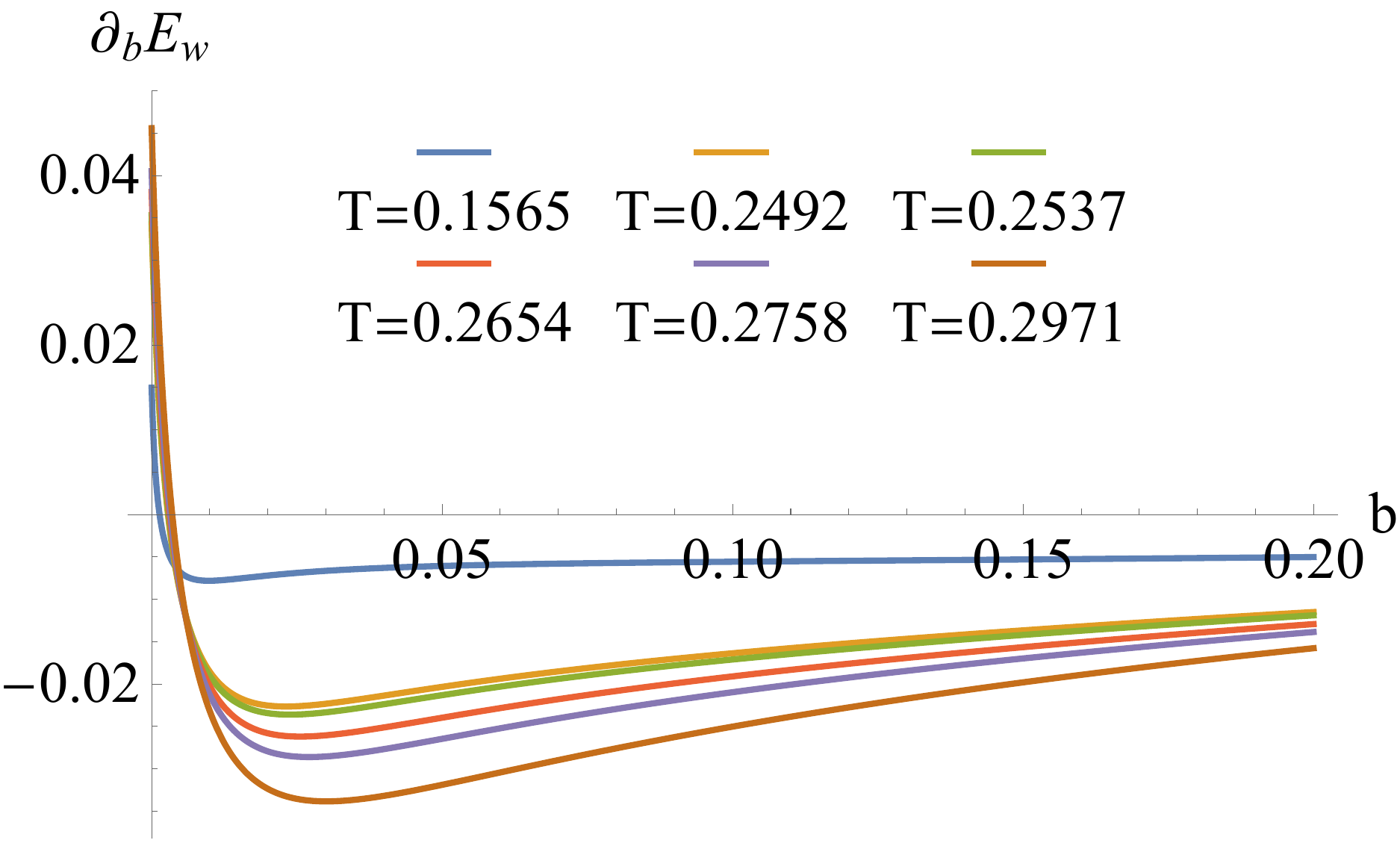}
  \caption{EWCS vs $T$. This plot is obtained at $(a,p,c) =(0.1, 0.05, 0.06925)$. When $b$ is relatively large, the $E_W$ converges to certain fixed values. For $T=0.2971$ it can first increase, and later decreases, and after that increases with $b$. Therefore, for very small $b$ the EWCS increases with $b$, irrespective of the values of the $T$ and the configurations.}
  \label{fig:eopvsdb}
\end{figure}

We show the EWCS vs $b$ in Fig. \ref{fig:eopvsdb}, from which we can find that the EWCS can show very delicate behaviors. The EWCS increases with $b$ at first in a very narrow range of $b$, however, it starts to decrease with $b$ when $b$ is relatively large and monotonically decreases with $b$. This is in sharp contrast to the behavior of the HEE and MI, that only shows monotonical behaviors (see Fig. \ref{fig:heetest} and Fig. \ref{fig:mi}). In addition, the $E_W$ changes slower with $b$ than that with $T$. The typical change is of order $10^{-4}$ and $10^{-3}$, respectively. This delicate behavior can be captured precisely because the precision of our numerical methods can be up to $10^{-7}$.

Notice that the background is an AdS-Schwarzschild solution when $ b $ is 0, meanwhile, its electromagnetic field is non-zero. At this point, the charge transport behavior of the system is significantly different from that of the genuine AdS-Schwarzschild system. Moreover, since its geometry is still AdS-Schwarzschild, the entanglement-related geometric quantities will be decoupled from the charge transport. As $ b $ gradually increases, the background geometry will receive back reactions from the Maxwell field. At this time, the entanglement-related geometry starts to couple with the charge transports. Therefore, $ b $ can play a role in measuring the relationship between entanglement and transport when $b$ is small. As we have pointed out, the EWCS increases with $ b $ when $ b $ is very small, i.e., when the coupling has just occurred. And when $ b $ increases further, the EWCS gradually shows a decreasing behavior. Notice that simpler geometric quantities such as HEE, and MI only show a very flat monotonic behavior. This indicates that EWCS, as a mixed-state entanglement, captures very different properties from HEE and MI.

To understand the above behavior more clearly, we implement the following analytical treatments. For small values of $b$, we can expand the expression of the $E_W$ \eqref{eq:zpara} integral with respect to $b$ as,
\begin{equation}\label{eq:dewdbsmall}
    E_W = \int_\Sigma \left(\frac{\sqrt{\Xi(z)}}{z^2} + \frac{b \left(\Gamma \left(\frac{1}{4}\right)+8 \Gamma \left(\frac{5}{4}\right)\right)}{2 \sqrt{3 \pi T \Xi(z) } \left(z^2+z+1\right) \Gamma \left(\frac{1}{4}\right) } + O(b^2)\right)dz,
  \end{equation}
where $\Xi(z) \equiv  {x'(z)}^2 + \frac{1}{\left(1-z^3\right)}$ and the second term shows us that $\frac{dE_W}{db}>0$ for small values of $b$. This explains the ubiquitous existence of the monotonically increasing behavior of $E_W$ vs $b$ for small values of $b$. From the holographic dual picture, it means that when the Maxwell field starts to turn on from the BI case, the $E_W$ is increased. However, when further increasing $b$ we find that $E_W$ reaches local maximums and starts to decrease. When $b$ is large, it can be expected that the background system approaches the AdS-RN, a fixed background geometry. Therefore, the $E_W$ will starts to converge to some fixed value. 
  For sufficiently large values of $b$, it is possible to derive the analytical expression for $E_W$ through expansions. The expression is given by Equation \ref{eq:dewdblarge},
  \begin{equation}\label{eq:dewdblarge}
    E_W = \int_\Sigma \left(\frac{\sqrt{\Theta(z)}}{z^2} - \frac{ z Q^4 \left(z^4+z^3+z^2+z+1\right)}{40 b^2 (1-z) \left(-Q^2 z^3+z^2+z+1\right)^2 \sqrt{\Theta (z)}} + O\left(\frac{1}{b^3}\right) \right) d z,
  \end{equation}
  where $\Theta (z)$ is defined as,
  $$\Theta (z) \equiv x'(z)^2 + \frac{1}{(1-z) \left(-Q^2 z^3+z^2+z+1\right)} > 0.$$
  The second term in \eqref{eq:dewdblarge} reveals that $E_W$ decreases as $b$ becomes larger. Consequently, since $E_W$ increases for small values of $b$ and decreases for large values of $b$, there must exist at least one critical value of $b$ at which $E_W$ reaches its maximum for a continuous function. This observation aligns with the numerical results and provides an explanation for the non-monotonic behavior of $E_W$ with respect to $b$.

  In the context of gravity, the non-monotonic behavior of EWCS versus $b$ can be interpreted by its increasing tendencies at smaller $b$ values and decreasing tendencies at larger ones. What separates EWCS from the typical monotonic behavior of HEE and MI is its location within the bulk. This positioning frees it from the overpowering influence of thermal effects. Furthermore, EWCS is unique due to its second-order minimization process, compared to HEE and MI that are associated with first-order minimization. This allows EWCS to capture different degrees of freedom in the quantum system. On the flip side, in the dual field theory, mixed state entanglement, such as entanglement of purification, also involves the second-order minimization. This difference also provides an explanation for the distinct behavior observed.

\begin{figure}
  \centering
  \includegraphics[width=0.5\textwidth]{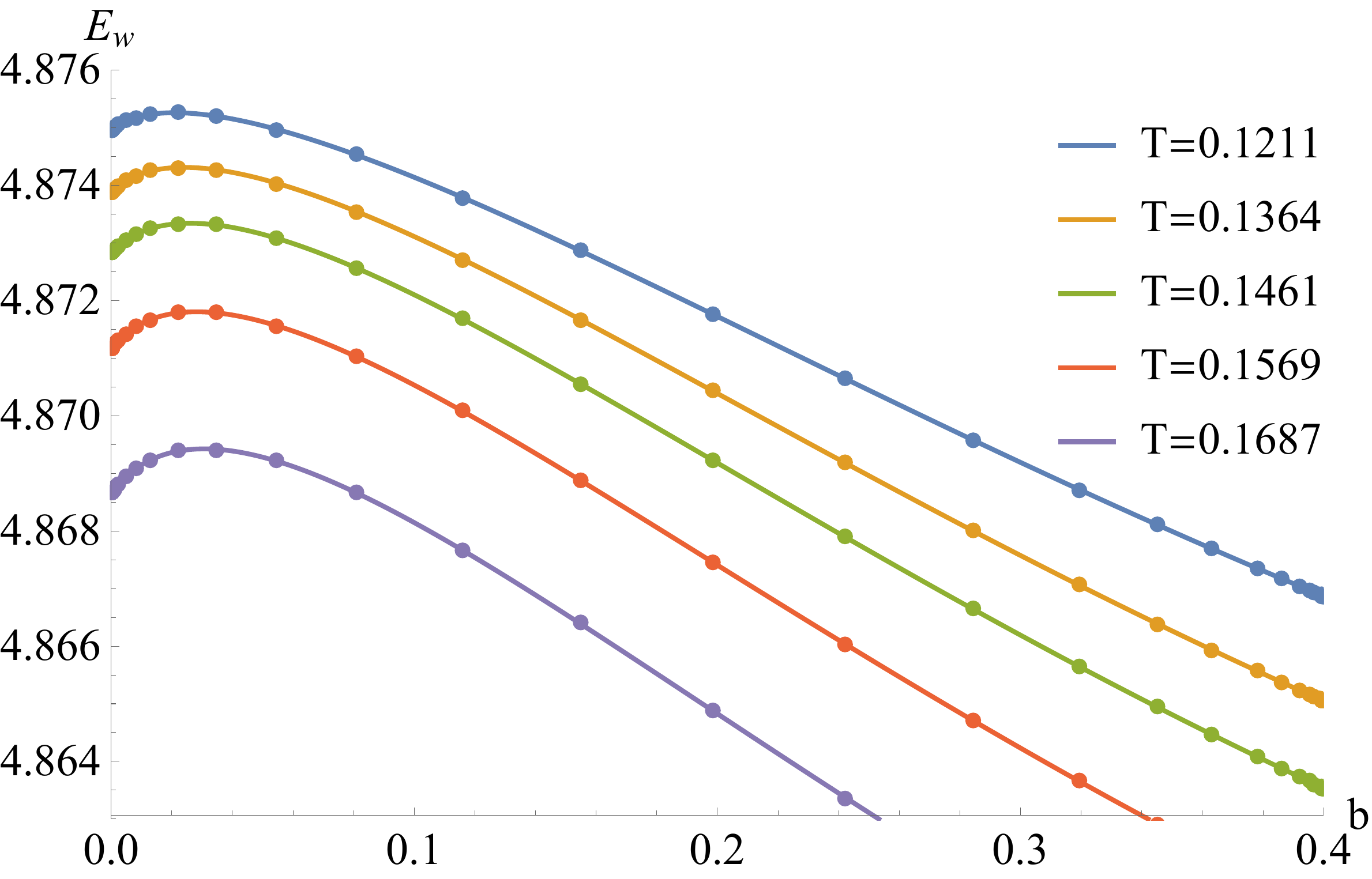}
  \caption{The EWCS vs $b$ for a larger configuration $(a,p,c) =(0.5, 0.2, 0.3875)$.}
  \label{fig:eopvsb4}
\end{figure}

Next, we show the EWCS in larger configurations in Fig. \ref{fig:eopvsb4}.
As seen in Fig. \ref{fig:eopvsb4}, the non-monotonicity of $E_W$ with $b$ becomes more pronounced as the width of the configuration increases. This means that the non-monotonicity exists over a wider interval. The reason for this is that when the width is relatively small, the minimum surface and the minimal cross-section are only slightly different from the properties of AdS. However, as the width increases, they deviate more significantly from AdS.

\begin{figure}
  \centering
  \includegraphics[width=0.45\textwidth]{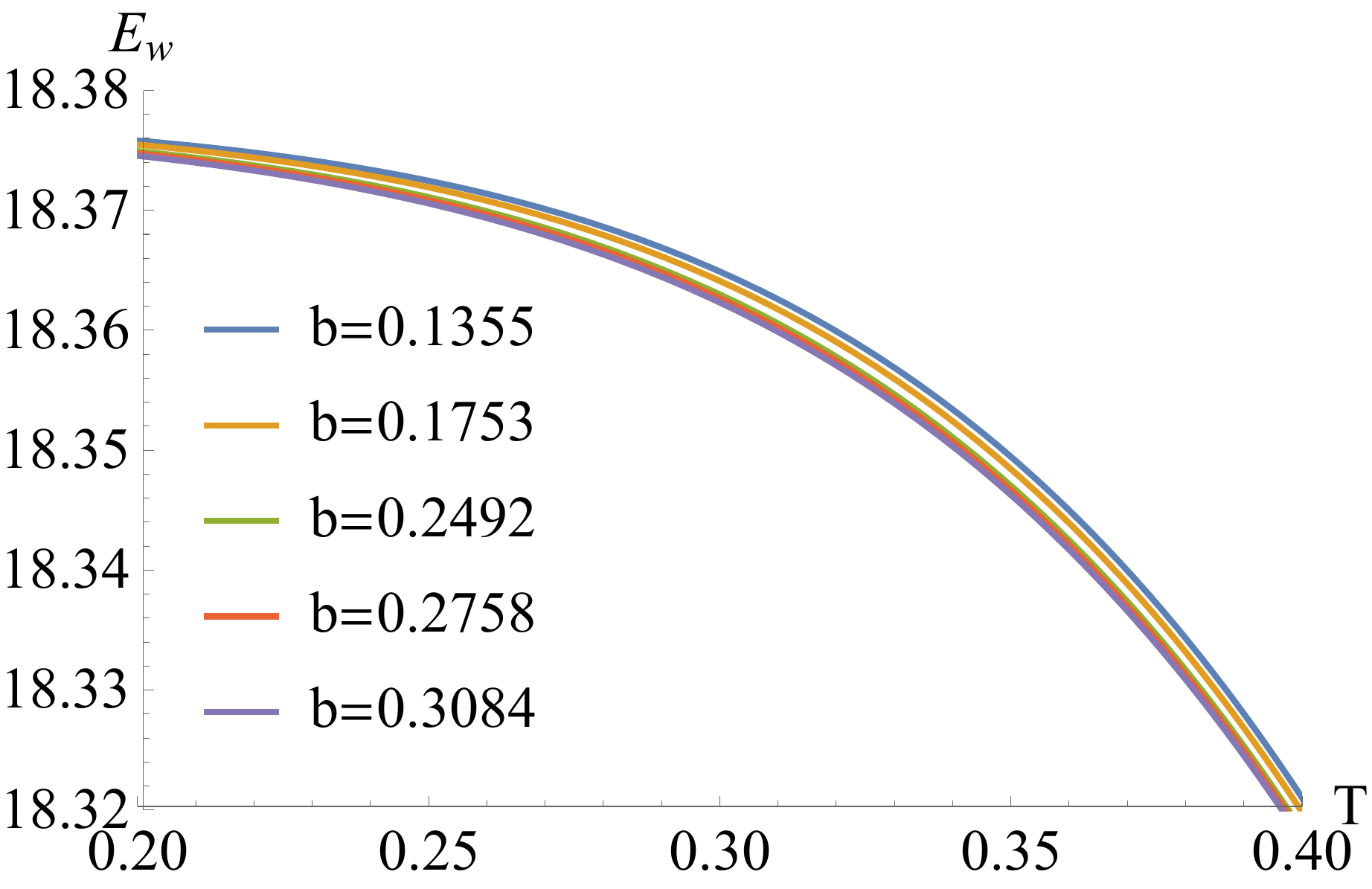}
  \includegraphics[width=0.45\textwidth]{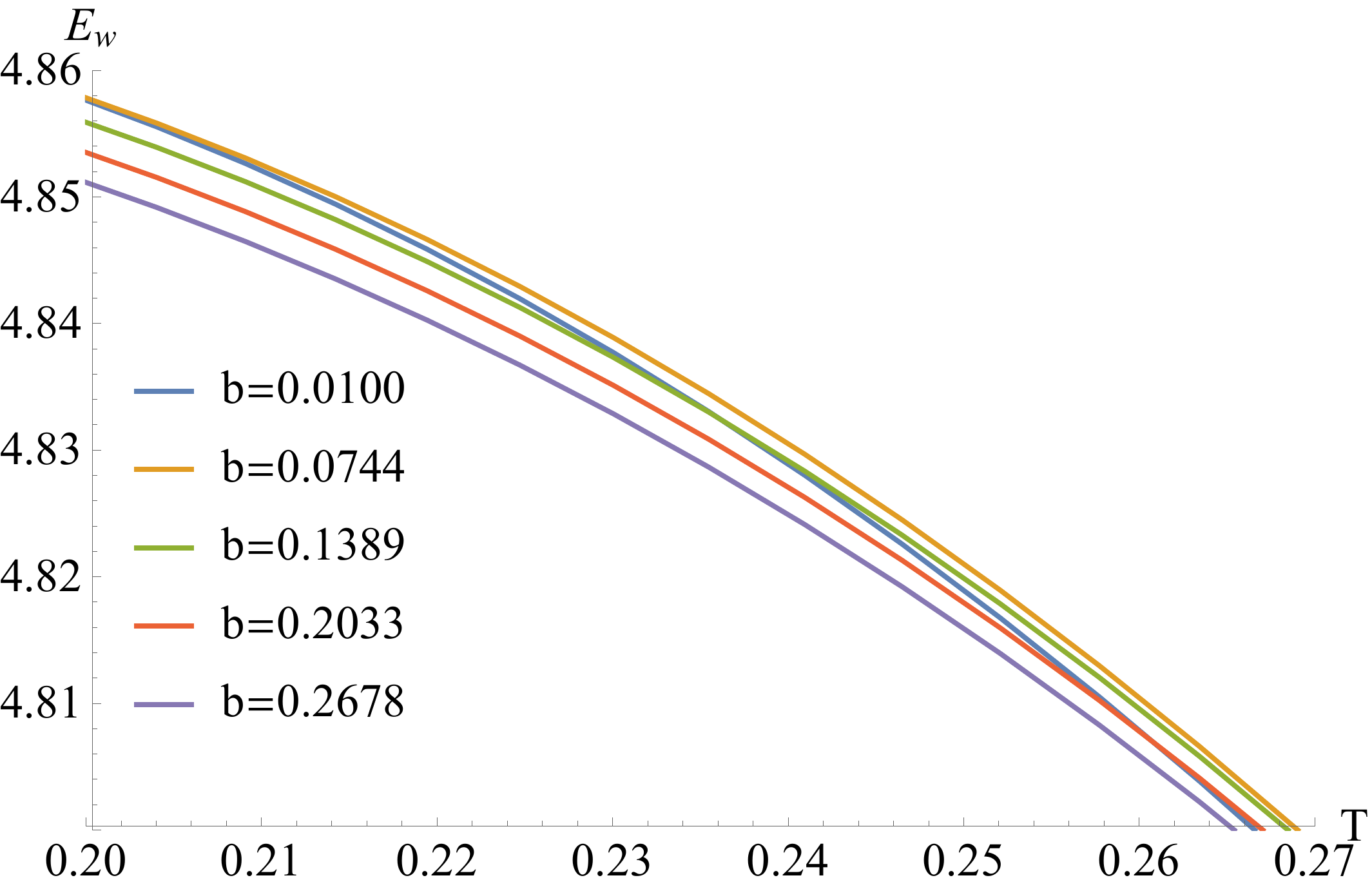}
  \caption{EWCS vs $T$. The left plot is obtained at $(a,p,c) =(0.1, 0.05, 0.06925)$; while the right plot is obtained for a larger configuration $(a,p,c) =(0.5, 0.2, 0.3875)$.}
  \label{fig:eopvst2}
\end{figure}

Next, we examine the behavior of EWCS with temperature. When the configuration is relatively small in BI systems, EWCS decreases monotonically with temperature, as shown in the left plot of Fig. \ref{fig:eopvst2}. It is worth noting that the non-monotonic behavior of EWCS at extremely small temperatures has been studied in \cite{Liu:2019qje} for AdS-RN systems. Additionally, we illustrate the behavior of EWCS with temperature for larger configurations in the right plot of Fig. \ref{fig:eopvst2}, which also shows that EWCS decreases monotonically with temperature. Although the monotonic decreasing behaviors are similar, the EWCS curves for small configurations differ from those for large configurations. By comparing the two plots in Fig. \ref{fig:eopvst2}, crossovers of the EWCS curves with temperature can be observed in the larger configuration, which reflects the non-monotonic behavior of EWCS with $b$. These findings suggest that the behavior of EWCS is generally monotonically decreasing with temperature, and this behavior is consistent with that of MI.

\begin{figure}
  \centering
  \includegraphics[width=0.7\textwidth]{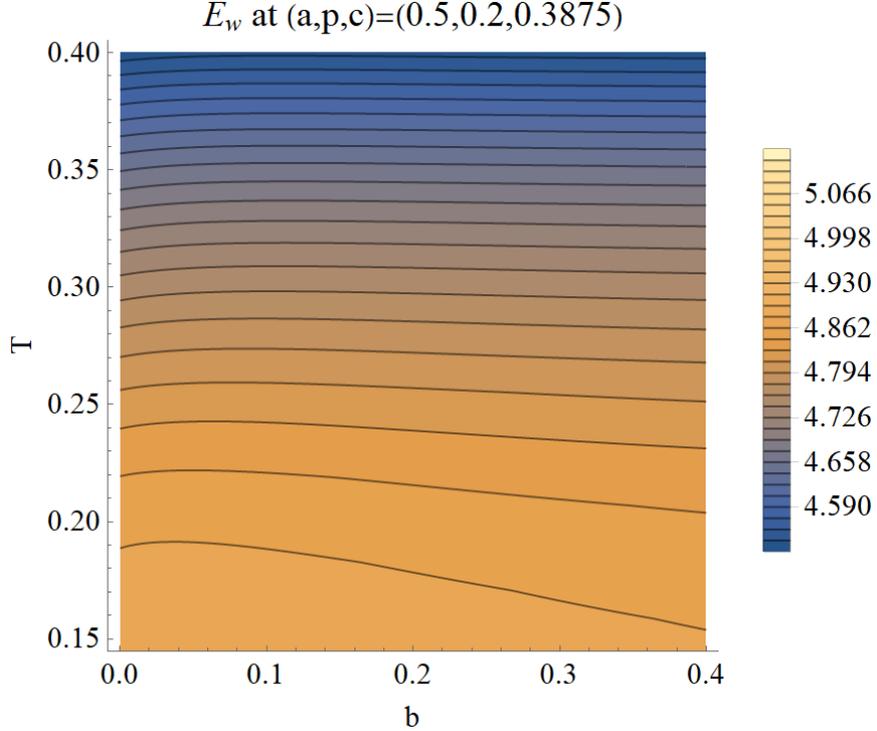}
  \caption{The EWCS vs $b$ for a larger configuration $(a,p,c) =(0.5, 0.2, 0.3875)$.}
  \label{fig:eopcontours1}
\end{figure}

In order to more clearly demonstrate the relationship between the EWCS and variables $b$ and $T$, a contour plot of EWCS as a function of $b$ and $T$ is presented in Fig. \ref{fig:eopcontours1}. This plot illustrates the non-monotonic nature of EWCS with respect to $b$ and the monotonic decrease of EWCS as $T$ increases. 

  The non-monotonic dependence of the EWCS on the BI factor $b$ that we have demonstrated is expected to be universal for BI-like nonlinear electromagnetism. To test this universality beyond the specific holographic model studied here, we will examine two additional BI-like theories. Observing similar non-monotonicity will confirm the crucial role of the BI term in generating this entanglement behavior, independently of other holographic dual details. Studying additional models will also elucidate the interplay between nonlinear electromagnetic effects and entanglement.

\section{Born-Infeld with Axions and Born-Infeld with massive gravity}\label{sec:biaxmg}

We aim to extend our investigation of holographic information measures by exploring two additional modifications of Born-Infeld (BI) theory. Specifically, we will examine BI theory coupled to axions and BI theory with massive gravity terms.

In the case of BI theory with axions, previous research focused on analyzing holographic transport in a model that incorporates the Born-Infeld term and axion fields $\phi_I = \alpha x_I$, where $I=1,2$ and $\alpha$ is a constant. The action for this model is given by \cite{Wu:2018zdc},
\begin{equation}
  S = \int d^4x \sqrt{-g} \left(R + \frac{6}{L^2} - \frac{1}{2}(\partial\phi_I)^2 + \frac { b ^ { 2 } } { 4 \pi G } \left( 1 - \sqrt { 1 + \frac { 2 \mathcal{F} } { b ^ { 2 } } } \right) \right),
\end{equation}
In this study, the DC and AC conductivities were analytically computed, revealing a temperature-dependent DC conductivity. The analysis also observed a transition from coherent to incoherent transport, which depended on parameters such as $b$.

For BI theory with massive gravity, the action considered includes the Einstein-Hilbert, Born-Infeld, and massive gravity terms \cite{Hendi:2015hoa},
\begin{equation}
  S = \int d^4 x \sqrt{-g} \left(R - 2\Lambda + m^2 \sum_{i=1}^4 c_iU_i(g,f) + \frac { b ^ { 2 } } { 4 \pi G } \left( 1 - \sqrt { 1 + \frac { 2 \mathcal{F} } { b ^ { 2 } } } \right) \right),
\end{equation}
Here, the $U_i$ represent symmetric polynomials of the eigenvalues of the $d\times d$ matrix $\mathcal{K}^\mu_\nu = \sqrt{g^{\mu\alpha}f_{\alpha\nu}}$, defined as follows:
\begin{align}
  U_1 & = [\mathcal{K}],                                                                                                              \\
  U_2 & = [\mathcal{K}]^2 - [\mathcal{K}^2],                                                                                          \\
  U_3 & = [\mathcal{K}]^3 - 3[\mathcal{K}][\mathcal{K}^2] + 2[\mathcal{K}^3],                                                         \\
  U_4 & = [\mathcal{K}]^4 - 6[\mathcal{K}]^2[\mathcal{K}^2] + 8[\mathcal{K}][\mathcal{K}^3] + 3[\mathcal{K}^2]^2 - 6[\mathcal{K}^4].
\end{align}
The authors obtained black hole solutions and conducted a comprehensive analysis of their thermodynamic stability and phase transitions. Notably, they found that properties such as the heat capacity $C_Q$ and phase structure were influenced by the nonlinear Born-Infeld coupling parameter $b$ and the massive gravity terms.

\begin{figure}[h]
  \centering 
  \includegraphics[width=0.45\textwidth]{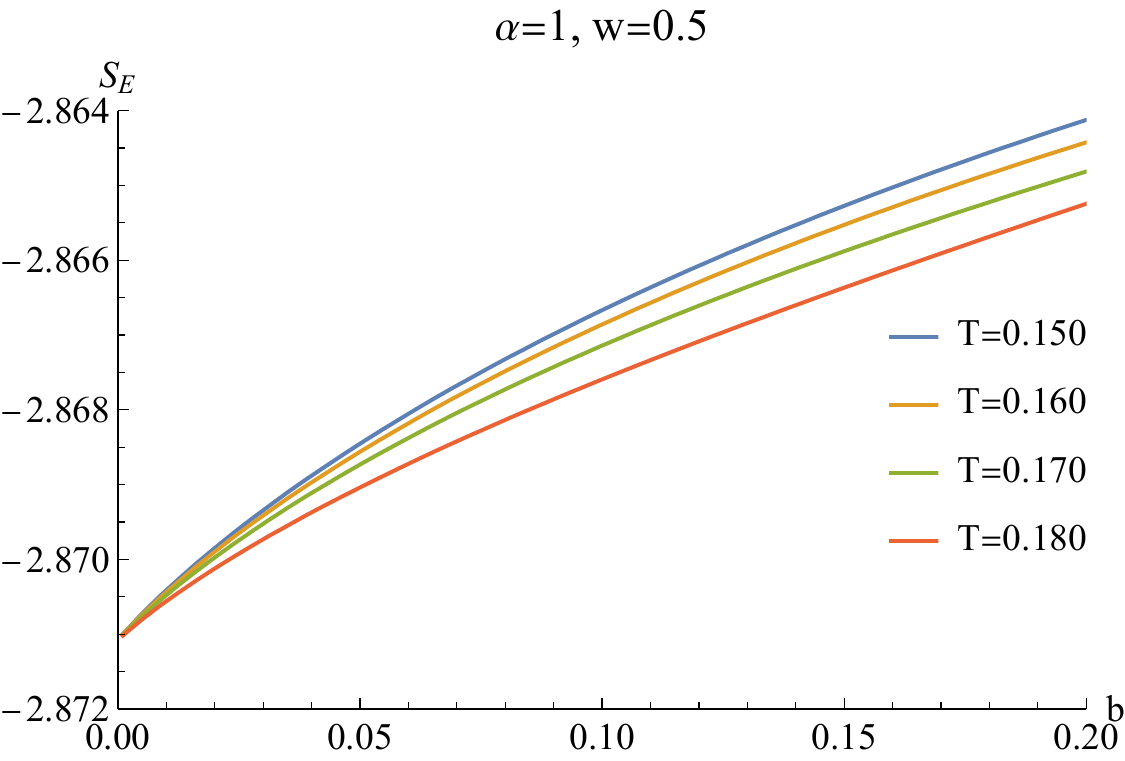}
  \includegraphics[width=0.45\textwidth]{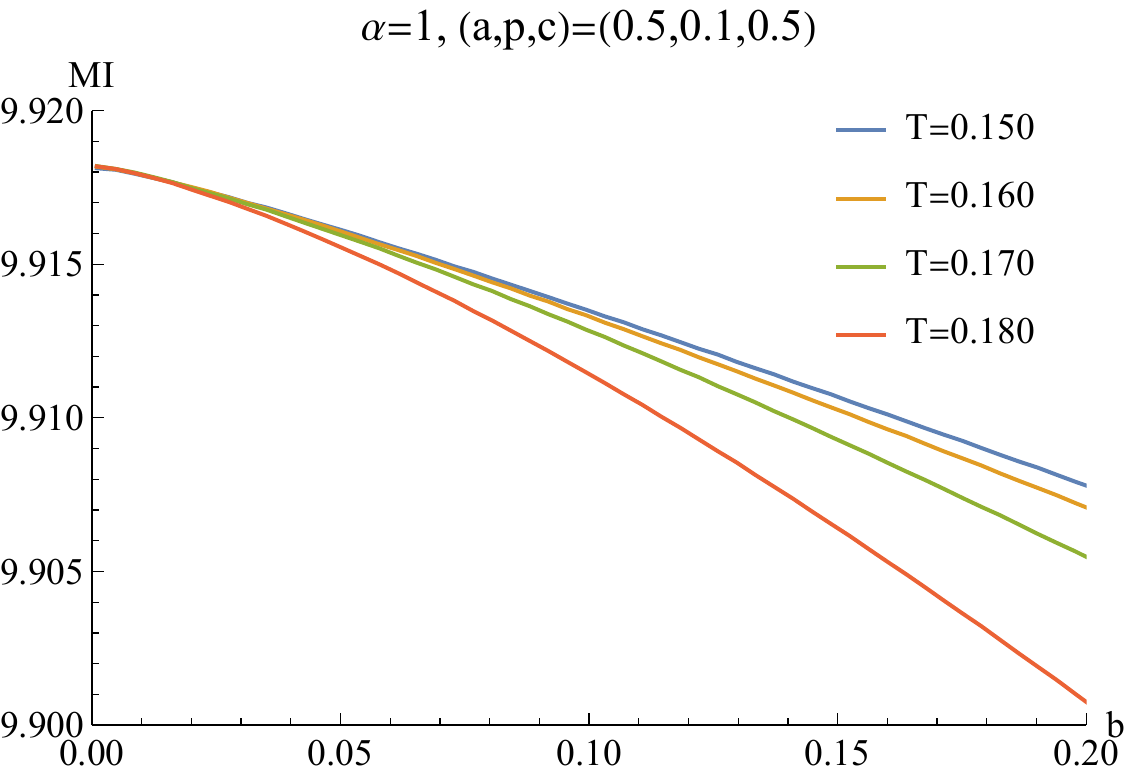}
  \caption{Left plot: HEE $S_E$ vs $b$ at $w=0.5$ and $\alpha=1$. Right plot: MI vs $b$ at $(a,p,c)=(0.5,1,0.5)$ and $\alpha=1$.}
  \label{fig:bihee}
\end{figure}

\begin{figure}[h]
  \centering
  \includegraphics[width=0.45\textwidth]{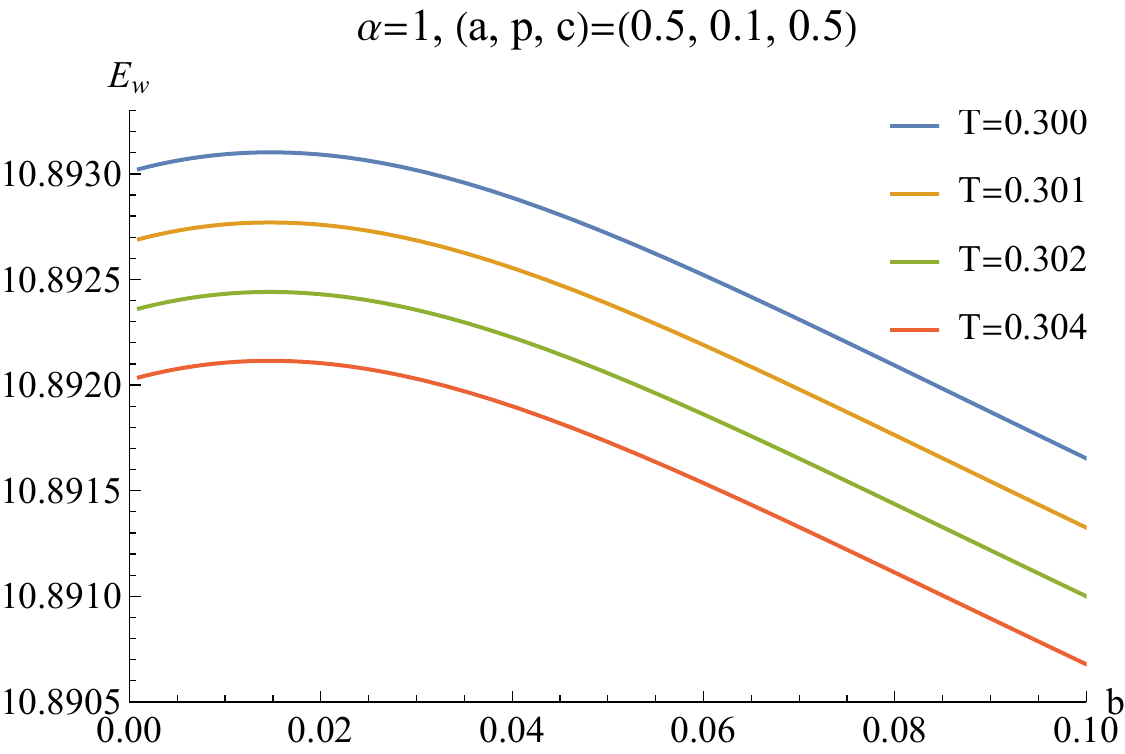}
  \includegraphics[width=0.45\textwidth]{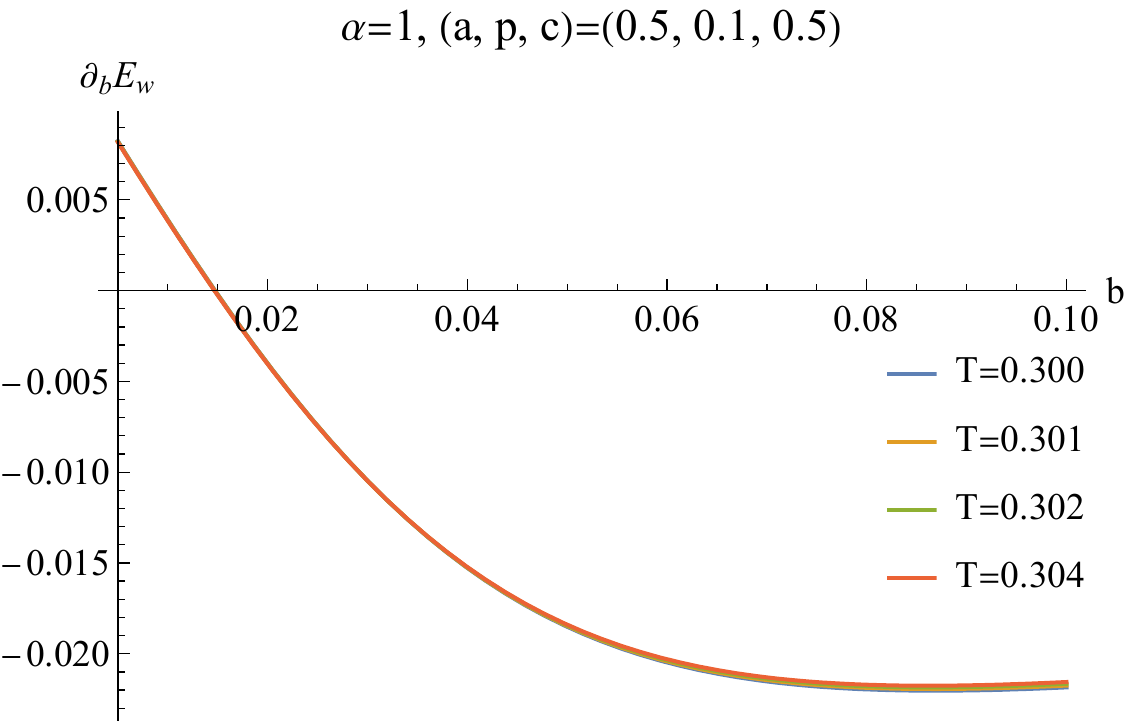}
  \caption{Configuration of the EWCS $(a,p,c)=(0.5,0.1,0.5)$. Left plot: EWCS $E_w$ versus the BI factor $b$ at different temperatures. Right plot: First derivatives of EWCS $E_w$ with respect to the BI factor $b$.}
  \label{fig:enbihee}
\end{figure}

First, we present the results for the BI-axion model, where we examine the behavior of holographic entanglement entropy (HEE) and mutual information (MI) with respect to the parameter $b$. The corresponding figures are shown in Fig. \ref{fig:bihee}.
It is evident that the HEE monotonically increases with $b$, while the MI decreases. Additionally, Fig. \ref{fig:enbihee} demonstrates the non-monotonic behavior of the entanglement wedge cross-section (EWCS).
The EWCS also exhibits non-monotonic behavior with respect to $b$. These observations in the BI-axion model align with those observed in the pure BI model.

\begin{figure}[h]
  \centering 
  \includegraphics[width=0.45\textwidth]{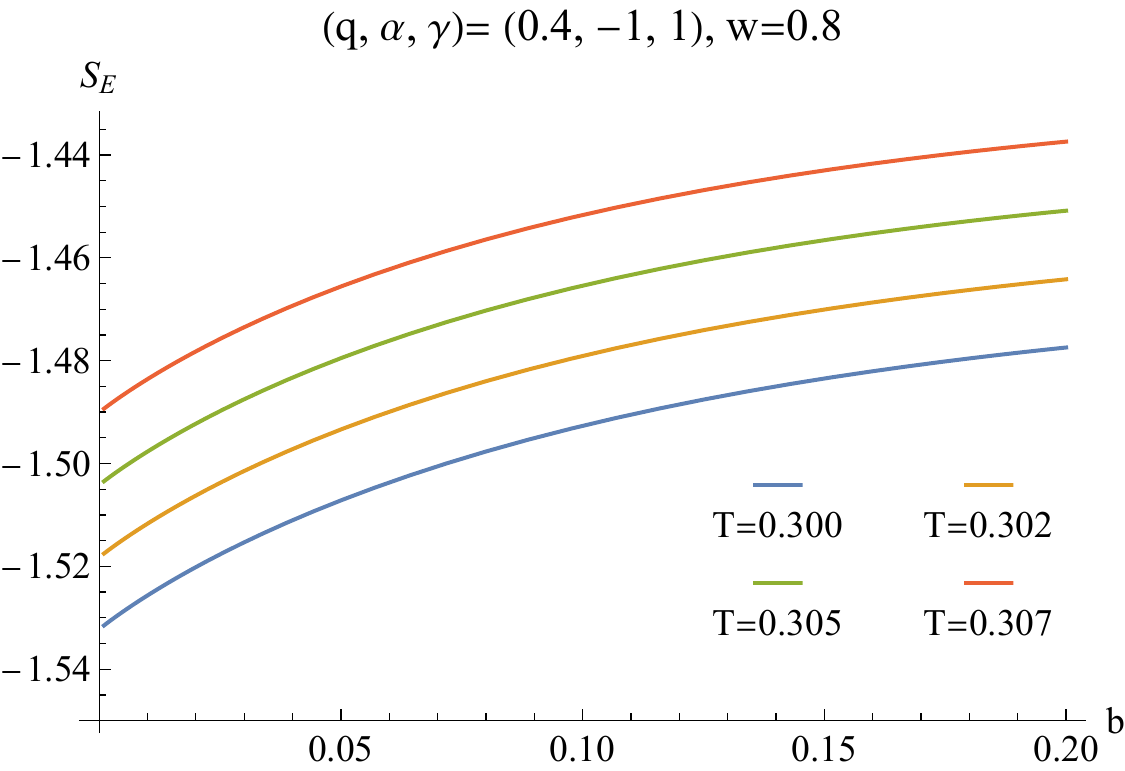}
  \includegraphics[width=0.45\textwidth]{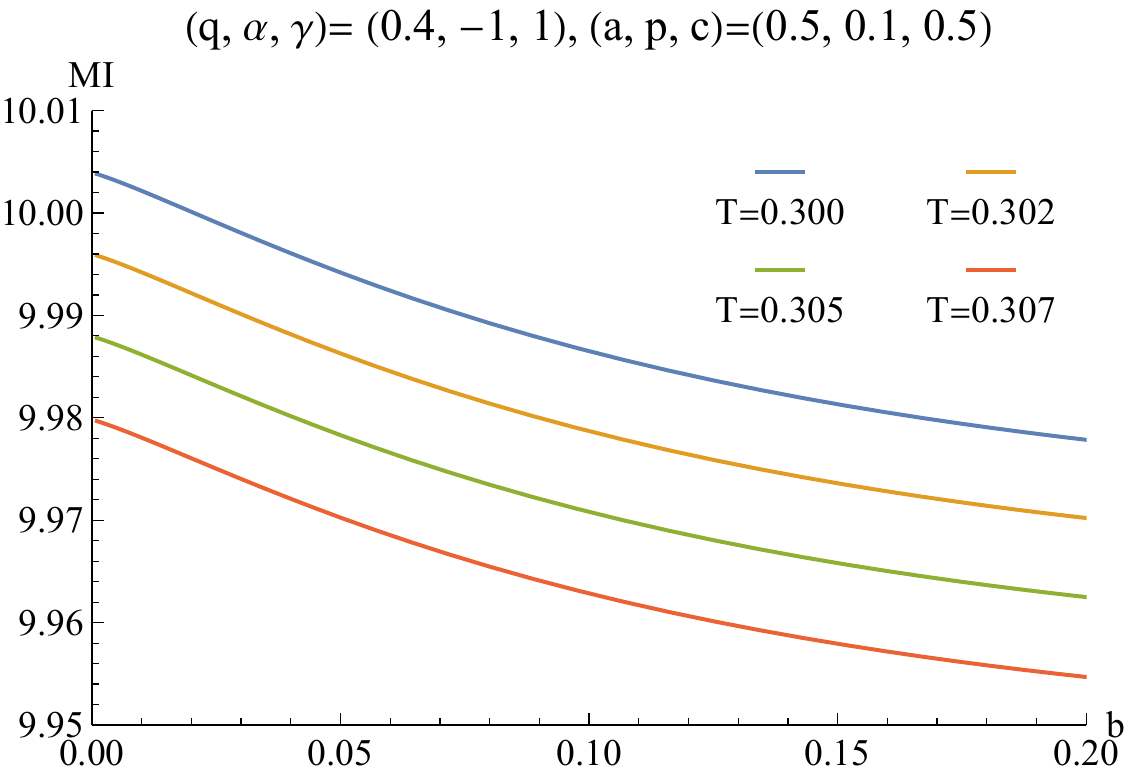}
  \caption{Left plot: HEE $S_E$ versus $b$ at $w=0.8$. Right plot: MI vs $b$ at $(a,p,c)=0.5,1,0.5$.}
  \label{fig:enbimgee}
\end{figure}

\begin{figure}
  \centering 
  \includegraphics[width=0.45\textwidth]{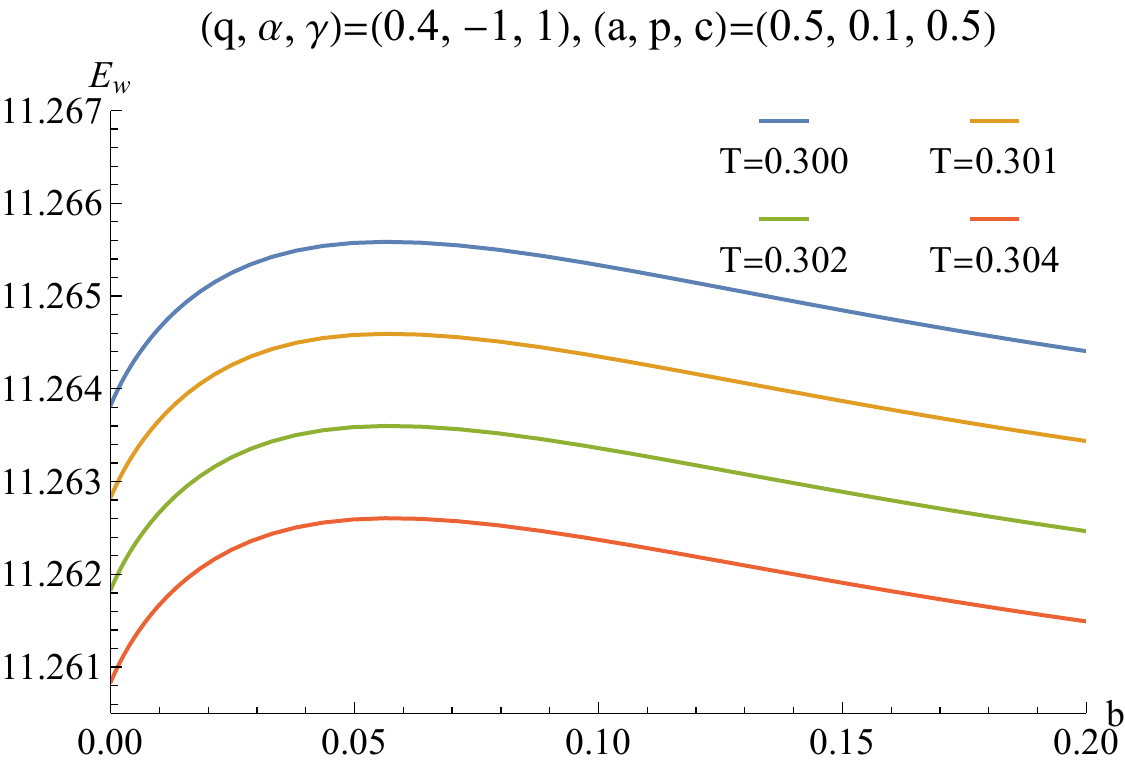}
  \includegraphics[width=0.45\textwidth]{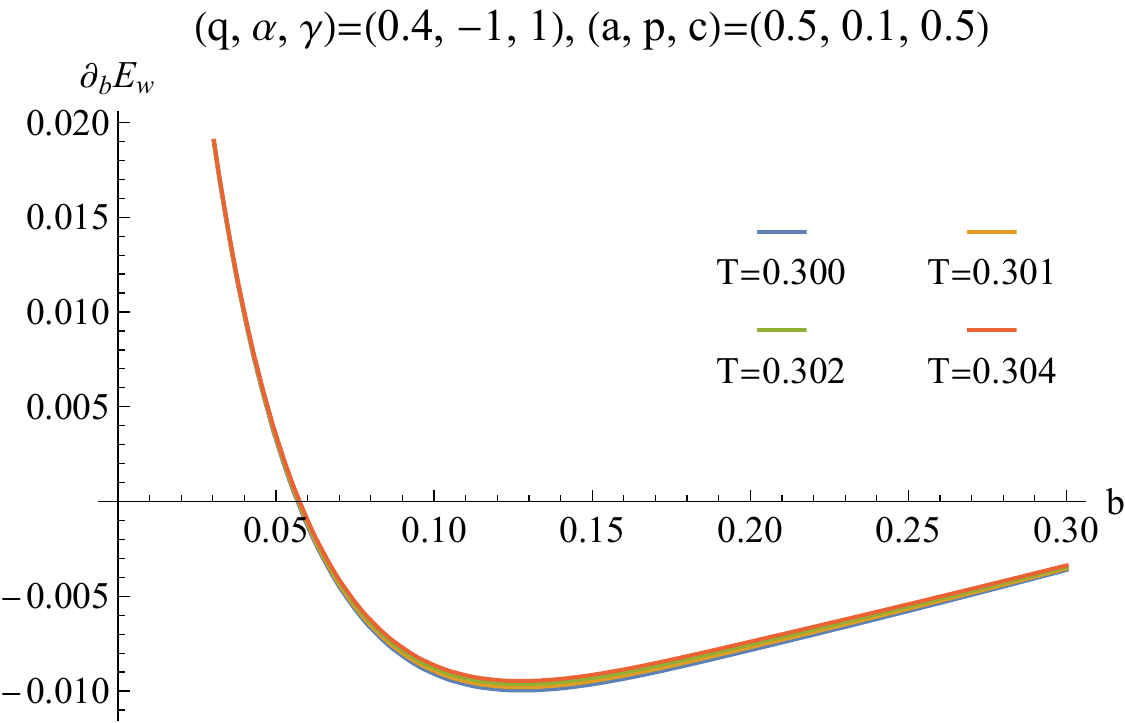}
  \caption{Entanglement wedge cross-section (EWCS) $E_w$ versus the BI factor $b$ with $(a,p,c)=(0.5,0.1,0.5)$. Left plot: $E_w$ versus $b$ at different temperatures. Right plot: First derivative of $E_w$ with respect to $b$, demonstrating the non-monotonic behavior of EWCS.}\label{fig:biaxewcs}
\end{figure}

Next, we investigate the BI model with massive gravity. Fig. \ref{fig:enbimgee} illustrates the behavior of HEE and MI with respect to the parameter $b$.
Similar to the BI-axion model, the HEE monotonically increases with $b$, while the MI decreases. Furthermore, Fig. \ref{fig:biaxewcs} presents the non-monotonic behavior of the EWCS in the EN-BI massive gravity model.
Again, the EWCS exhibits non-monotonicity with respect to $b$. The phenomena observed in the BI-axion model are consistent with those observed in the pure BI model.

Based on the aforementioned observations in both the BI-axion model and the BI model with massive gravity, we propose that the Born-Infeld term governs the monotonic behavior of HEE, MI, and EWCS with respect to $b$, as all three models exhibit qualitatively similar results. This suggests the universality of the phenomenon across BI-like theories, highlighting the significance of the Born-Infeld term in modeling the coupling between entanglement and transport in holographic theories. Consequently, further exploration of BI gravity is warranted to gain insights into strongly correlated systems.

\section{Discussion}\label{sec:discuss}

In this paper, we study the behavior of HEE, MI, and the mixed-state entanglement measure EWCS in the BI model. Our results show that HEE increases monotonically with both $b$ and $T$, while MI decreases monotonically with both $b$ and $T$. Interestingly, the behavior of EWCS with respect to $b$ shows a non-monotonic trend. Specifically, when $b$ is small, EWCS increases with $b$, but it begins to decrease as $b$ increases further. In contrast, EWCS exhibits a consistent monotonically decreasing trend with $T$.

Note that when $b$ is small, $b$ serves as a measure of the coupling between the entanglement-related quantities and the charge transport of the system. Based on this observation, we conjecture that increasing the coupling between the entanglement-related quantities and the transport properties can enhance the EWCS of the system. This coupling between transport behaviors and entanglement is also a topic of significant interest in condensed matter theory, as seen in previous studies on nanowires \cite{nanowire:2011}, plasmonics \cite{plasmonics:2013,plasmon:2004}, and plasmons \cite{plasom:2002}.

  The HEE increases consistently with temperature $T$ and the BI factor $b$ in the gravity perspective. This is due to the expansion of the black hole horizon radius $r_h$, which brings the HEE minimum surface closer to the horizon. As a result, the HEE behavior is primarily influenced by the thermal entropy from the black hole. From the viewpoint of the dual CFT, this monotonic relationship between HEE and $ T $, as well as $ b $, arises because thermal effects dominate the entanglement entropy. Similarly, the MI decreases continuously with $ T $ and $ b $, as it is defined based on the HEE. On the CFT side, increasing $ T $ and $ b $ introduce more thermal noise, which disrupts entanglement in the boundary theory.

  In contrast, the non-monotonic behavior of the EWCS is characterized by an initial increase for small values of $b$ followed by a subsequent decrease for large values of $b$. This unique pattern emerges because the EWCS relies on the cross-section within the bulk interior, independent of the horizon's influence. From a dual CFT perspective, the EWCS serves as a probe for mixed state entanglement, which effectively avoids thermal noise. Moreover, measures of mixed state entanglement, such as the entanglement of purification, involve a second-order minimization process rather than a first-order one like HEE. This second-order minimization enables the EWCS to capture new quantum information that is overlooked by the thermal-dominated first-order measures like HEE or MI.

  To test the universality of our results, we examined two additional BI-like holographic models. In both models, we observed monotonic HEE and MI together with non-monotonic EWCS dependence on the BI factor $b$. This alignment across the models suggests the non-monotonic EWCS phenomenon originates specifically from the Born-Infeld term, independently of other holographic details. Our observations highlight the crucial role of the BI term in governing the interplay between nonlinear electromagnetic effects and entanglement. The universality motivates further study of BI gravity to elucidate strongly correlated systems.

Next, we point out the issues that deserve further investigation. To begin, we can examine other BI-like theories, such as the BI theory with lattices or superconductivity, to further test if the non-monotonic behavior observed in this paper is general. Furthermore, we can examine the effect of more general nonlinear EM field theories on the entanglement-related physical quantities of the system, such as the more general nonlinear EM fields \cite{Guo:2017bru,Baggioli:2016oju}. We are working on these directions.

\section*{Acknowledgments}

Peng Liu would like to thank Yun-Ha Zha for her kind encouragement during this work. Zhe Yang would like to express appreciation to Feng-Ying Deng. This work is supported by the Natural Science Foundation of China under Grant No. 11805083, 11905083, 12005077, 12147209, 12375055 the Science and Technology Planning Project of Guangzhou (202201010655) and Guangdong Basic and Applied Basic Research Foundation (2021A1515012374). J.-P.W. is also supported by Top Talent Support Program from Yangzhou University.


\begin{thebibliography}{99}
  
  \bibitem{Osterloh:2002na}
  A. Osterloh, L. Amico, G. Falci, R. Fazio,
  ``Scaling of Entanglement close to a Quantum Phase Transitions''
  Nature {\bf 416}, 608 (2002) [arXiv:0202029 [quant-ph]]
  
  \bibitem{Amico:2007ag}
  L. Amico, R. Fazio, A. Osterloh and V. Vedral,
  ``Entanglement in many-body systems''
  Rev.Mod.Phys. {\bf 80}, 517 (2008)
  [arXiv:0703044 [quant-ph]]
  
  \bibitem{Wen:2006topo}
  Levin, Michael, and Xiao-Gang Wen.
  ``Detecting topological order in a ground state wave function''. Physical review letters 96.11 (2006): 110405.
  
  \bibitem{Kitaev:2006topo}
  Kitaev, Alexei, and John Preskill.
  ``Topological entanglement entropy''. Physical review letters 96.11 (2006): 110404.
  
  \bibitem{Ryu:2006bv}
  S.~Ryu and T.~Takayanagi,
  ``Holographic derivation of entanglement entropy from AdS/CFT,''
  Phys.\ Rev.\ Lett.\  {\bf 96}, 181602 (2006)
  [hep-th/0603001].
  
  \bibitem{Hubeny:2007xt}
  V.~E.~Hubeny, M.~Rangamani and T.~Takayanagi,
  ``A Covariant holographic entanglement entropy proposal,''
  JHEP {\bf 0707}, 062 (2007)
  [arXiv:0705.0016 [hep-th]].
  
  \bibitem{Lewkowycz:2013nqa}
  A.~Lewkowycz and J.~Maldacena,
  ``Generalized gravitational entropy,''
  JHEP {\bf 1308}, 090 (2013)
  [arXiv:1304.4926 [hep-th]].
  
  \bibitem{Dong:2016hjy}
  X.~Dong, A.~Lewkowycz and M.~Rangamani,
  ``Deriving covariant holographic entanglement,''
  JHEP {\bf 1611}, 028 (2016)
  [arXiv:1607.07506 [hep-th]].
  
  \bibitem{vidal:2002}
  Vidal, G. and Werner, R.F., 2002.
  ``A computable measure of entanglement'',
  Physical Review A, 65(3), p.032314. quant-ph:0102117
  
  \bibitem{Horodecki:2009review}
  Horodecki, R., Horodecki, P., Horodecki, M.,  Horodecki, K. (2009).
  ``Quantum entanglement.''
  Reviews of modern physics, 81(2), 865.
  
  \bibitem{Nishioka:2006gr}
  T.~Nishioka and T.~Takayanagi,
  ``AdS Bubbles, Entropy and Closed String Tachyons,''
  JHEP {\bf 0701}, 090 (2007)
  [hep-th/0611035].
  
  \bibitem{Klebanov:2007ws}
  I.~R.~Klebanov, D.~Kutasov and A.~Murugan,
  ``Entanglement as a probe of confinement,''
  Nucl.\ Phys.\ B {\bf 796}, 274 (2008)
  [arXiv:0709.2140 [hep-th]].
  
  \bibitem{Pakman:2008ui}
  A.~Pakman and A.~Parnachev,
  ``Topological Entanglement Entropy and Holography,''
  JHEP {\bf 0807}, 097 (2008)
  [arXiv:0805.1891 [hep-th]].
  
  \bibitem{Zhang:2016rcm}
  S.~J.~Zhang,
  ``Holographic entanglement entropy close to crossover/phase transition in strongly coupled systems,''
  Nucl.\ Phys.\ B {\bf 916}, 304 (2017)
  [arXiv:1608.03072 [hep-th]].
  
  \bibitem{Zeng:2016fsb}
  X.~X.~Zeng and L.~F.~Li,
  ``Holographic Phase Transition Probed by Nonlocal Observables,''
  Adv.\ High Energy Phys.\  {\bf 2016}, 6153435 (2016)
  [arXiv:1609.06535 [hep-th]].
  
  \bibitem{Dong:2016fnf}
  X.~Dong,
  ``The Gravity Dual of Renyi Entropy,''
  Nature Commun.\  {\bf 7}, 12472 (2016)
  [arXiv:1601.06788 [hep-th]].
  
  \bibitem{Shenker:2013pqa}
  S.~H.~Shenker and D.~Stanford,
  ``Black holes and the butterfly effect,''
  JHEP {\bf 1403}, 067 (2014)
  [arXiv:1306.0622 [hep-th]].
  
  \bibitem{Sekino:2008he}
  Y.~Sekino and L.~Susskind,
  ``Fast Scramblers,''
  JHEP {\bf 0810}, 065 (2008)
  [arXiv:0808.2096 [hep-th]].
  
  \bibitem{Maldacena:2015waa}
  J.~Maldacena, S.~H.~Shenker and D.~Stanford,
  ``A bound on chaos,''
  JHEP {\bf 1608}, 106 (2016)
  [arXiv:1503.01409 [hep-th]].
  
  \bibitem{Donos:2012js}
  A.~Donos and S.~A.~Hartnoll,
  ``Metal-insulator transition in holography'',
  Nature Phys.\  {\bf 9}, 649 (2013)
  [arXiv:1212.2998].
  
  \bibitem{Blake:2016wvh}
  M.~Blake,
  ``Universal Charge Diffusion and the Butterfly Effect in Holographic Theories,''
  Phys.\ Rev.\ Lett.\  {\bf 117}, no. 9, 091601 (2016)
  [arXiv:1603.08510 [hep-th]].
  
  \bibitem{Blake:2016sud}
  M.~Blake,
  ``Universal Diffusion in Incoherent Black Holes,''
  Phys.\ Rev.\ D {\bf 94}, no. 8, 086014 (2016)
  [arXiv:1604.01754 [hep-th]].
  
  \bibitem{Ling:2016ibq}
  Y.~Ling, P.~Liu and J.~P.~Wu,
  ``Holographic Butterfly Effect at Quantum Critical Points,''
  JHEP {\bf 1710}, 025 (2017)
  [arXiv:1610.02669 [hep-th]].
  
  \bibitem{Ling:2016wuy}
  Y.~Ling, P.~Liu and J.~P.~Wu,
  ``Note on the butterfly effect in holographic superconductor models,''
  Phys.\ Lett.\ B {\bf 768}, 288 (2017)
  [arXiv:1610.07146 [hep-th]].
  
  \bibitem{Wu:2017mdl}
  S.~F.~Wu, B.~Wang, X.~H.~Ge and Y.~Tian,
  ``Collective diffusion and quantum chaos in holography,''
  Phys.\ Rev.\ D {\bf 97}, no. 10, 106018 (2018)
  [arXiv:1702.08803 [hep-th]].
  
  \bibitem{Liu:2019npm}
  P.~Liu, C.~Niu and J.~P.~Wu,
  ``The Effect of Anisotropy on Holographic Entanglement Entropy and Mutual Information,''
  Phys.\ Lett.\ B {\bf 796}, 155 (2019)
  [arXiv:1905.06808 [hep-th]].
  
  \bibitem{Brown:2015lvg}
  A.~R.~Brown, D.~A.~Roberts, L.~Susskind, B.~Swingle and Y.~Zhao,
  ``Complexity, action, and black holes'',
  Phys.\ Rev.\ D {\bf 93}, no. 8, 086006 (2016)
  [arXiv:1512.04993 [hep-th]].
  
  \bibitem{Brown:2015bva}
  A.~R.~Brown, D.~A.~Roberts, L.~Susskind, B.~Swingle and Y.~Zhao,
  ``Holographic Complexity Equals Bulk Action?''
  Phys.\ Rev.\ Lett.\  {\bf 116}, no. 19, 191301 (2016)
  [arXiv:1509.07876 [hep-th]].
  
  \bibitem{Chapman:2016hwi}
  S.~Chapman, H.~Marrochio and R.~C.~Myers,
  ``Complexity of Formation in Holography'',
  JHEP {\bf 1701}, 062 (2017)
  [arXiv:1610.08063 [hep-th]].
  
  \bibitem{Ling:2018xpc}
  Y.~Ling, Y.~Liu and C.~Y.~Zhang,
  ``Holographic Subregion Complexity in Einstein-Born-Infeld theory,''
  Eur.\ Phys.\ J.\ C {\bf 79}, no. 3, 194 (2019)
  [arXiv:1808.10169 [hep-th]].
  
  \bibitem{Chen:2018mcc}
  B.~Chen, W.~M.~Li, R.~Q.~Yang, C.~Y.~Zhang and S.~J.~Zhang,
  ``Holographic subregion complexity under a thermal quench,''
  JHEP {\bf 1807}, 034 (2018)
  [arXiv:1803.06680 [hep-th]].
  
  \bibitem{Yang:2019gce}
  R.~Q.~Yang, H.~S.~Jeong, C.~Niu and K.~Y.~Kim,
  ``Complexity of Holographic Superconductors,''
  JHEP {\bf 1904}, 146 (2019)
  [arXiv:1902.07586 [hep-th]].
  
  \bibitem{Ling:2019ien}
  Y.~Ling, Y.~Liu, C.~Niu, Y.~Xiao and C.~Y.~Zhang,
  ``Holographic Subregion Complexity in General Vaidya Geometry,''
  JHEP {\bf 1911}, 039 (2019)
  [arXiv:1908.06432 [hep-th]].
  
  \bibitem{Takayanagi:2017knl}
  T.~Takayanagi and K.~Umemoto,
  ``Holographic entanglement wedge cross-section,''
  arXiv:1708.09393 [hep-th].
  
  \bibitem{Nguyen:2017yqw}
  P.~Nguyen, T.~Devakul, M.~G.~Halbasch, M.~P.~Zaletel and B.~Swingle,
  ``entanglement wedge cross-section: from spin chains to holography,''
  JHEP {\bf 1801}, 098 (2018)
  [arXiv:1709.07424 [hep-th]].
  
  \bibitem{Chen:2021bjt}
  C.~Y.~Chen, W.~Xiong, C.~Niu, C.~Y.~Zhang and P.~Liu,
  ``{\it Entanglement wedge minimum cross-section for holographic aether gravity,}''
  JHEP \textbf{08} (2022), 123
  [arXiv:2109.03733 [hep-th]].
  
  \bibitem{Cheng:2021hbw}
  F.~J.~Cheng, Z.~Yang, C.~Niu, C.~Y.~Zhang and P.~Liu,
  ``{\it Entanglement Wedge Minimum Cross-Section in Holographic Axion Gravity Theories,}''
  [arXiv:2109.03696 [hep-th]].
  
  \bibitem{Fu:2020oep}
  G.~Fu, P.~Liu, H.~Gong, X.~M.~Kuang and J.~P.~Wu,
  ``{\it Holographic informational properties for a specific Einstein-Maxwell-dilaton gravity theory,}''
  Phys. Rev. D \textbf{104} (2021) no.2, 026016
  [arXiv:2007.06001 [hep-th]].
  
  \bibitem{Gong:2020pse}
  H.~Gong, P.~Liu, G.~Fu, X.~M.~Kuang and J.~P.~Wu,
  ``{\it Informational properties of holographic Lifshitz field theory,}''
  Chin. Phys. C \textbf{45} (2021) no.6, 6
  [arXiv:2009.00450 [hep-th]].
  
  \bibitem{Ling:2021vxe}
  Y.~Ling, P.~Liu, Y.~Liu, C.~Niu, Z.~Y.~Xian and C.~Y.~Zhang,
  ``{\it Reflected entropy in double holography,}''
  JHEP \textbf{02} (2022), 037
  [arXiv:2109.09243 [hep-th]].
  
  \bibitem{Liu:2020blk}
  P.~Liu and J.~P.~Wu,
  ``{\it Mixed state entanglement and thermal phase transitions,}''
  Phys. Rev. D \textbf{104} (2021) no.4, 046017
  [arXiv:2009.01529 [hep-th]].
  
  \bibitem{Liu:2021rks}
  P.~Liu, C.~Niu, Z.~J.~Shi and C.~Y.~Zhang,
  ``{\it Entanglement wedge minimum cross-section in holographic massive gravity theory,}''
  JHEP \textbf{08} (2021), 113
  [arXiv:2104.08070 [hep-th]].
  
  \bibitem{Li:2021rff}
  Y.~Z.~Li, C.~Y.~Zhang and X.~M.~Kuang,
  ``Entanglement wedge cross-section with Gauss-Bonnet corrections and thermal quench,''
  Sci. China Phys. Mech. Astron. \textbf{64} (2021) no.12, 120413
  [arXiv:2102.12171 [hep-th]].
  
  \bibitem{Huang:2019zph}
  Y.~f.~Huang, Z.~j.~Shi, C.~Niu, C.~y.~Zhang and P.~Liu,
  ``Mixed-state Entanglement for Holographic Axion Model,''
  Eur. Phys. J. C \textbf{80} (2020) no.5, 426
  [arXiv:1911.10977 [hep-th]].
  
  \bibitem{Kundu:2013eba}
  A.~Kundu and S.~Kundu,
  ``Steady-state Physics, Effective Temperature Dynamics in Holography,''
  Phys. Rev. D \textbf{91} (2015) no.4, 046004
  [arXiv:1307.6607 [hep-th]].
  
  \bibitem{Kundu:2019ull}
  A.~Kundu,
  ``Steady States, Thermal Physics, and Holography,''
  Adv. High Energy Phys. \textbf{2019} (2019), 2635917
  
  \bibitem{Karch:2009zz}
  A.~Karch, D.~T.~Son and A.~O.~Starinets,
  ``Holographic Quantum Liquid,''
  Phys. Rev. Lett. \textbf{102} (2009), 051602
  
  \bibitem{Baggioli:2016oju}
  M.~Baggioli and O.~Pujolas,
  ``On Effective Holographic Mott Insulators,''
  JHEP \textbf{12}, 107 (2016)
  [arXiv:1604.08915 [hep-th]].
  
  \bibitem{Kiritsis:2016cpm}
  E.~Kiritsis and L.~Li,
  ``Quantum Criticality and DBI Magneto-resistance,''
  J. Phys. A \textbf{50} (2017) no.11, 115402
  [arXiv:1608.02598 [cond-mat.str-el]].
  
  \bibitem{Cremonini:2017qwq}
  S.~Cremonini, A.~Hoover and L.~Li,
  ``Backreacted DBI Magnetotransport with Momentum Dissipation,''
  JHEP \textbf{10}, 133 (2017)
  [arXiv:1707.01505 [hep-th]].
  
  \bibitem{Hayers:2014zz}
  I. M. Hayers, N. P. Breznay, T. Helm, P. Moll, M. Wartenbe, R. D. McDonald, A.
  Shekhter, J. G. Analytis, {\it Magnetoresistance near a quantum critical point},
  [ArXiv:1412.6484][cond-mat.str-el];
  
  \bibitem{Hayes:2016}
  I. M. Hayes, R. D. McDonald, N. P. Breznay, T. Helm, P. J. W. Moll, M. Wartenbe,
  A. Shekhter, J. G. Analytis, {\it Scaling between magnetic field and temperature in the high-temperature superconductor BaFe$_2$(As$_{1-x}$P$_x$)$_2$}, Nature Physics (2016).
  
  \bibitem{Wu:2018zdc}
  J.~P.~Wu, X.~M.~Kuang and Z.~Zhou,
  ``Holographic transports from Born\textendash{}Infeld electrodynamics with momentum dissipation,''
  Eur. Phys. J. C \textbf{78}, no.11, 900 (2018)
  [arXiv:1805.07904 [hep-th]].
  
  \bibitem{Bakhtiarizadeh:2020kav}
  H.~R.~Bakhtiarizadeh and G.~Jafari,
  ``Holographic complexity of Born\textendash{}Infeld gravity,''
  Eur. Phys. J. C \textbf{80}, no.3, 208 (2020)
  [arXiv:2002.09974 [hep-th]].
  
  \bibitem{Mejia-Monasterio:2007}
  C. Mejia-Monasterio and H. Wichterich,
    ``Heat transport in quantum spin chains: Stochastic baths vs quantum trajectories,''
    The European Physical Journal Special Topics,
    \textbf{151}, 113-125 (2007)
  
  
  \bibitem{Hasan:2010}
  M. Z. Hasan and C. L. Kane,
    ``Colloquium: Topological insulators,''
    Rev. Mod. Phys. \textbf{82}, 3045 (2010)
  
  
  \bibitem{Stoermer:1999}
  Horst L. Stormer,
    ``Nobel Lecture: The fractional quantum Hall effect,''
    Rev. Mod. Phys. \textbf{71}, 875 (1999)
  
  
  \bibitem{Nandkishore:2015}
  R. Nandkishore and D. A. Huse,
    ``Many-Body Localization and Thermalization in Quantum Statistical Mechanics,''
    Annual Review of Condensed Matter Physics,
    \textbf{6}: 15-38 (Volume publication date March 2015)
  
  
  \bibitem{nanowire:2011}
  Chen, Guang-Yin, et al. ``Surface Plasmons in a Metal Nanowire Coupled to Colloidal Quantum Dots: Scattering Properties and Quantum Entanglement.''
  Physical Review B, vol. 84, no. 4, July 2011.
  
  \bibitem{plasom:2002}
  Altewischer, E., van Exter, M. P. \& Woerdman, ``J. P. Plasmon-assisted transmission of entangled photons.'' 
  Nature 418, 304-306 (2002).
  
  \bibitem{plasmonics:2013}
  Tame, M., McEnery, K., \"Ozdemir, . et al. 
  ``Quantum plasmonics." 
  Nature Phys 9, 329-340 (2013).
  
  \bibitem{Cai:2004eh}
  R.~G.~Cai, D.~W.~Pang and A.~Wang,
  ``Born-Infeld black holes in (A)dS spaces,''
  Phys. Rev. D \textbf{70} (2004), 124034
  [arXiv:hep-th/0410158 [hep-th]].
  
  \bibitem{plasmon:2004}
  Moreno, E., Garc\'ia, F. J., Erni, D., Ignacio Cirac, J. \& Mart\'in-Moreno, L.
  ``Theory of plasmon-assisted transmission of entangled photons." Phys. Rev.
  Lett. 92, 236801 (2004).
  
  \bibitem{Chuang:2002book}
  Nielsen, Michael A., and Isaac Chuang.
  ``Quantum computation and quantum information."
  (2002): 558-559.
  
  \bibitem{Kudler-Flam:2018qjo}
  J.~Kudler-Flam and S.~Ryu,
  ``Entanglement negativity and minimal entanglement wedge cross sections in holographic theories,''
  Phys. Rev. D \textbf{99} (2019) no.10, 106014
  [arXiv:1808.00446 [hep-th]].
  
  \bibitem{Kusuki:2019zsp}
  Y.~Kusuki, J.~Kudler-Flam and S.~Ryu,
  ``Derivation of holographic negativity in AdS$_3$/CFT$_2$,''
  Phys. Rev. Lett. \textbf{123} (2019) no.13, 131603
  [arXiv:1907.07824 [hep-th]].
  
  \bibitem{Dutta:2019gen}
  S.~Dutta and T.~Faulkner,
  ``A canonical purification for the entanglement wedge cross-section,''
  JHEP \textbf{03} (2021), 178
  [arXiv:1905.00577 [hep-th]].
  
  \bibitem{Bao:2017nhh}
  N.~Bao and I.~F.~Halpern,
  ``Holographic Inequalities and entanglement wedge cross-section,''
  JHEP {\bf 1803}, 006 (2018)
  [arXiv:1710.07643 [hep-th]].
  
  \bibitem{Ling:2019tbi}
  Y.~Ling, Y.~Liu and Z.~Y.~Xian,
  ``Entanglement Entropy of Annulus in Holographic Thermalization,''
  arXiv:1911.03716 [hep-th].
  
  \bibitem{Boyd:2001}
  John P Boyd. {\it Chebyshev and Fourier spectral methods}. Courier Corporation, 2001.
  
  \bibitem{Liu:2019qje}
  P.~Liu, Y.~Ling, C.~Niu and J.~P.~Wu,
  ``entanglement wedge cross-section in Holographic Systems,''
  JHEP \textbf{09} (2019), 071
  [arXiv:1902.02243 [hep-th]].
  
  \bibitem{Hendi:2015hoa}
  S.~H.~Hendi, B.~Eslam Panah and S.~Panahiyan,
  ``Einstein-Born-Infeld-Massive Gravity: adS-Black Hole Solutions and their Thermodynamical properties,''
  JHEP \textbf{11} (2015), 157
  [arXiv:1508.01311 [hep-th]].
  
  \bibitem{Guo:2017bru}
  X.~Guo, P.~Wang and H.~Yang,
  ``Membrane Paradigm and Holographic DC Conductivity for Nonlinear Electrodynamics,''
  Phys. Rev. D \textbf{98} (2018) no.2, 026021
  [arXiv:1711.03298 [hep-th]].
  
\end{thebibliography}
\end{document}